\documentclass{aa}  

\usepackage{graphicx}   % Including figure files
\usepackage{amsmath}    % Advanced maths commands
\usepackage{comment}
\usepackage{siunitx}
\usepackage{booktabs}
\usepackage{float}
\DeclareSIUnit\Jansky{Jy}
\DeclareSIUnit{\erg}{erg}
\usepackage{txfonts}
\usepackage{hyperref}
\usepackage[nameinlink,capitalise]{cleveref}
\crefname{section}{Sect.}{Sects.}
\def\subinrm#1{\sb{\mathrm{#1}}}
{\catcode`\_=13 \global\let_=\subinrm}
\mathcode`_="8000
\def\upsubscripts{\catcode`\_=12 } 
\newcommand{\fracsi}[1]{\si[per-mode=fraction]{#1}}
\begin{document} 

\upsubscripts

   \title{Searching for radio emission from radio quiet magnetars with MeerKAT}

   \author{Marlon L. Bause\inst{1},
           Kamalpreet Kaur\inst{1},
           Isabella Rammala-Zitha\inst{1}
          \and
          Laura G. Spitler\inst{1}
          }

   \institute{Max Planck Institut für Radioastronomie,
   Auf dem H\"ugel 69, 53121 Bonn, Germany\\
              \email{mbause@mpifr-bonn.mpg.de}
             }
    \authorrunning{M. L. Bause et al.}
   \date{Received 07 August, 2025; accepted XXXX, 2025}

% \abstract{}{}{}{}{} 
% 5 {} token are mandatory
 
  \abstract
   {Magnetars are neutron stars that occupy the extreme end of the neutron star population, with magnetic field strengths of more than $10^{12}$~\si{G}. They have been proposed as one of the most likely progenitor models for the phenomenon of energetic, ms-duration, extragalactic radio bursts (FRBs), which has been increased even further due to the FRB-like bursts emitted from the galactic Magnetar SGR 1935+2154.  However, only a low fraction of the magnetars (six in total) has been detected in the radio regime and thus most magnetars are radio quiet.}
   {We conducted regular observations of 13 radio quiet magnetars to probe the long term radio quietness using the most sensitive telescope in the southern hemisphere: MeerKAT. These observations provide deep constraints on the radio emission of magnetars, relevant for the progenitor models of FRBs}
   {Given that MeerKAT is a interferometer, we probe the magnetars for radio emission in both imaging and time domain. We search in the time domain in a DM range of \SIrange{20}{10000}{pc \per cm^3} for single pulses using a TransientX based search pipeline (the FRB perspective) as well as from a pulsar perspective by folding the data using the X-ray ephemeris. On the other hand, we use the imaging domain to search for persistent radio emission in total intensity and circular polarisation as well as to create light curves using snapshot imaging having the long transient perspective as well.}
   {We find no radio emission in the time domain for any of the observed magnetars. Nevertheless, we are able to provide deep limits of the the mean flux density (\SIrange{52}{68}{\micro Jy}) and the single pulse fluence \SIrange{39}{52}{\milli \Jansky \milli \second}. From the image domain, we provide individual upper limits on the persistent radio radio emission and the light curve for the 13 magnetars. Additionally, an ultra long period transient and an additional magnetar happened to be in the imaging beam for which be provide lower limits as well.}
   {We provide an extensive series of deep upper limits in the time domain but also as a novelty limits from the imaging domain for the magnetars. As the current magnetar radio emission models are based on a few radio loud magnetars, we encourage monitoring of radio quiet magnetars independent of their X-ray flux with high cadence for further insights in their potential for emitting in the radio regime.}

   
   \maketitle
%
%-------------------------------------------------------------------

\section{Introduction}
    Magnetars are a sub-class of neutron stars, whose inferred dipole magnetic field strengths are the highest seen in neutron stars (\SIrange{e12}{e15}{G}). The term was originally introduced by \cite{duncan1992} and \cite{thomposon1993} as a unified model of Soft Gamma-ray Repeaters (SGRs) and Anormalous X-ray Pulsars (AXPs).
    The two main observable characteristic behaviours of magnetars are strong and frequent glitches and the high energy emission in the X-ray (and partially also gamma-ray).
    Glitches are sudden jumps in the spin period of the neutron star, which seem to be related to the interior of the neutron star as discussed by \citet{anderson1975}.
    The high energy emission consists of a persistent profile of pulsed emission with periods in the order of \SIrange{2}{12}{s} as well as transient emission in the form of bursts, which are often clustered in time but not necessarily follow the rotational period \citep{gogus1999} on ms to s scale as well as rare (giant) flare events \citet{kaspi2017}.
    Additionally, magnetars can undergo outbursts, in which the X-ray flux increases by a factor of 10 to 1000 and shows enhanced bursting activity.
    Thus, classifying a newly found neutron star, which is not showing the hallmark observables, as a magnetar can be challenging as only a inferred high magnetic field and the pulsed (X-ray) emission can be observed.
    %characteristics generally is of transient nature but differs between the quiescent state and the (post-) outburst state.
    %In the quiescent state, magnetars show , while in outburst many bursts, which are often clustered in time but not not necessarily %follow the rotational period \citep{gogus1999} on ms to s scale as well as rare (giant) flare events \citet{kaspi2017} can be observed.
    However, the nature of the transient X-ray emission is highly time variable and can change on time scales of weeks.
    The magnetar catalogue\footnote{\url{https://www.physics.mcgill.ca/~pulsar/magnetar/main.html}}\citep{olausen2014} includes currently 30 magnetars and magnetar candidates.
    
    Of these magnetars, six have been observed in radio: SGR 1935+2154 (for example \citet{bochenek2020, chimesgr19352020}), XTE J1810-197 (for example \citet{halpern2005}), Swift J1818.0-1607 (for example \citet{Karuppusamy2020}), SGR J1745-2900 (for example \citet{shannon2013}), PSR J1622-4950 (for example \citet{levin2010}) and 1E 1547.0-5408 (for example \citet{camilo2007}).
    The radio emission consists typically of short, ms-scale radio pulses as well as pulsed emission at the rotational period of the magnetar, often displayed as a pulse profile.
    However, the appearance of the profiles, as well as the single pulses, varies with time and can show sudden changes.
    For example in XTE J1810-197, as seen amongst others by \citet{levin2019} and \citep{bause2024}, the profile changes the number of components and the brightness of the single pulses changes significantly.
    Additionally, the radio emission is itself appears and disappears on a time scale of months to a few years \citep{camilo2016}.
    While it is typically assumed that radio emission follows after an outburst of a magnetar, \citet{caleb2022} and \citet{bause2024} have shown that the peak radio intensities of the magnetar XTE J1810+197 happen at a phase where the X-ray activity has left the outburst state and is back to its thermal emission.
    This indicates that this paradigm of searching for radio emission only after an X-ray outburst could be biased.
    Thus, motivating an X-ray outburst independent monitoring campaign of radio quiet magnetars.
   
    %Whether radio emission from a magnetar is expected or not, is thus unclear as there is no general  independent of the X-ray activity publicly available.
    From the theory side, there are attempts to constrain the parameters under which magnetars are capable of emitting in the radio regime.
    Assuming that the radio emission mechanism of magnetars, which is not well understood, follows similar conditions as the emissions mechanism of radio pulsars (that is a polar cap model), one can define the so-called death line(s) of spin period and its derivative under which the magnetic field is too strong, and thus quenches the radio emission as shown by \citet{kaiyou1993}.
    Depending on the configuration of the magnetic field assumed for the death line, several radio quiet magnetars lay in the region where radio emission is expected, while radio loud magnetars lay below the death line.
    Hence, these alone do not suffice to predict radio emission from a magnetar.
    
    Additionally, the so-called fundamental plane proposed by \citet{rea2012} is another approach to explain why only specific magnetars show radio emission by comparing the X-ray luminosity and the spin down luminosity (energy that is available from the spin down of the neutron star).
    If the X-ray luminosity is larger than the spin down luminosity, the energy source must come elsewhere, likely from the strong magnetic field. Under such conditions the magnetar clearly differs from radio pulsars, and no radio emission is expected.
    On the other hand, the magnetar would be powered by rotation if its spin down luminosity is higher than its X-ray luminosity, making it essentially a (radio) pulsar with a high magnetic field.
    
    The clearly radio loud magnetar XTE J1810-197 challenges both of these predictions. It is centrally in the death valley (the area in between the death line for a simple dipolar magnetic field and a complex magnetic field), indicating a more complex magnetic field structure.
    Additionally, its X-ray luminosity exceeds the spin down luminosity indicating that it is not powered by rotational energy but magnetic energy instead but still being radio loud.
    
    %Whether one might expect radio emission from a magnetar or not could be investigated using the period period derivative diagram. Following \textbf{referece}, the magnetospere of a magnetar becomes charge depleted if the magnetic field strength becomes to large.
    %Assuming a specific composition of the neutron star and a structure of the magnetic field (dipole, twists etc.) one can find death lines, which represents a combination of period and period derivative beyond which no radio emission is expected. \textbf{Fill plots}.
    %\textbf{Fill PLot}.
    %\textbf{Fill with some indications}
    %On the other hand, the beam of the magnetars which should show radio emission might be formed unlucky, so that the radio emission does not reach us.
    %This is referred to by the beaming fraction.

    Previous searches for radio emission typically targeted the source(s) sparsely or even just once after an X-ray outburst. This includes, for example, \cite{lorimer2000} for SGR 1900+14, while \cite{burgay_2006} and \cite{crawford_deep_2007} targeting four magnetars with one Parkes observations each.
    \cite{lazarus_constraining_2012} conducted a first campaign of targeting several targets in the search for radio emission, focusing on a target after an X-ray outburst with the Green Bank Telescope (GBT). Even the the Five-hundred-meter Aperture Spherical Telescope (FAST) has been used for searches in the radio emission of individual magnetars by \cite{lu2024}, \cite{bai2025} and \cite{xie2025}.

    Recently, magnetars have become the most actively followed progenitor models for fast radio bursts (FRBs), since a FRB-like burst has been detected from SGR 1935+2154 detected by \cite{bochenek2020} and \cite{chimesgr19352020}.
    This magnetar in particular has later been seen to emit sporadic single pulses as reported by \citet{kirsten2021} (detectable via single pulse (SP) searches) and faint pulsar like states \citet{zhu2023}. In this phase, the radio emission was visible in form of a folded profile as well as single pulses, that is different algorithms were used to find the radio emission (FFT and SP search).
    
    On the other hand, our understanding of under which conditions magnetars can emit radio emission is poor.
    Thus, gaining a better insight of it will also give crucial insights into eligibility of them to produce FRBs.
    Thus, we conducted a regular monitoring of twelve of the radio quiet magnetars as well as SGR 1935+2154 on a monthly basis with the MeerKAT telescope, which is a radio interferometer consisting of 64 dishes located in the Karoo semidesert in South Africa, which has the unique ability so collect beamforming and continuum data simultaneously. 
    Its high sensitivity and isolated location allow to find even the weakest emission and provide deep limits in the case of a non detection in both time and imaging domain.    

    This article is structured as follows: \Cref{sec:obs} describes the targets and observations, \Cref{sec:dr} describes our data reduction techniques, \Cref{sec:results} presents our results and \Cref{sec:discussion} and \Cref{sec:summary} related our findings to other works and summarise this work respectively.
    
\section{Observational strategy}\label{sec:obs}

\subsection{Targets}

From the list of magnetars that are published in the Magnetar catalogue, \citep{olausen2014}, we selected those that are visible by MeerKAT and within our Galaxy.
Furthermore, we also removed those only classified as magnetar candidate and those that have had radio emission previously detected with the exception of SGR 1935+2154 given its important role as a potential link to FRBs.
Hence, the following thirteen magnetars, listed in \Cref{tab:mag_properties}, were observed in our campaign:

\subsubsection{1E 1048.1-5937}
The magnetar 1E 1048.1-5937 was discovered by \cite{seward1986} in an X-ray survey with the High Energy Astronomy Observatory 2 (Einstein Observatory) in 1979.  Since its discovery, the magnetar has been active with several outbursts, flux enhancements and glitches (see \cite{archibalds2020} and references therein).
An infrared counterpart has been detected by \cite{wang2002} for this magnetar, which is also associated with a stellar wind bubble \citep{gaensler2005}. Previous searches for radio emission after outbursts have not revealed any radio emission \citep{2007ATelcamilo}.

\subsubsection{1E 1841-045}
\cite{vasisht1997} reported the detection of the magnetar 1E 1841-045, which is associated with the supernova remnant Kes 73 with observations of the Advanced Satellite for Cosmology and
Astrophysics as an anomalous X-ray pulsars (AXPs). Since its discovery, the magnetar has showed several periods of outbursts \citep{gavriil2002}, a behaviour that is known for soft gamma ray repeaters (SGRs). 
Thus, AXPs and SGRs were now observationally connected under the term magnetar. Additionally, this source went into outburst during our observational campaign as reported in \cite{younes2025}.

\subsubsection{1RXS J170849.0-400910}
The first detection of 1RXS J170849.0-400910 was reported by \cite{voges1999} in an all-sky survey by ROSAT (short for Röntgensatellit) with pulsations found by \cite{sugizaki1997}. Despite flaring activity reported by \cite{younes2020}, the magnetar has relatively stable X-ray flux levels \citep{dib2014}, as well as several glitches observed \citep{scholz2014}. Although a potential infrared counterpart \citep{durant2006} and an upper limit on mid-infrared \citep{wangb2007} have been found. This infrared emission is potentially related to interaction between the dust and the X-ray emission, this magnetar has only been clearly detected at high energies so far.

\subsubsection{3XMM J185246.6+003317}
\cite{zhou2014} and \cite{rea2014} both independently discovered 3XMM J185246.6+003317 using data from the X-ray Multi-Mirror Mission (XMM-Newton). Despite the X-ray emission, no low energy counter part  has been found. This only recently discovered magnetar has a comparably low magnetic field strength for magnetars (< \SI{4e13}{G}). Nevertheless, its change in spectrum and X-ray flux density classify it as a (transient) magnetar in a post outburst stage.

\subsubsection{CXOU J164710.2-455216}
CXOU J164710.2-455216 was discovered as a magnetar by \cite{muno2006} within the Westerlund I cluster with the Chandra X-ray telescope. As common for magnetars, it showed several outbursts \citep{woods2011, borghese2019}. Similar as 3XMM J185246.6+003317, this source is a magnetar with a rather low magnetic field of about \SI{4e13}{G} estimated by the timing solution of \citet{an2019}. A potential infrared counterpart as been found by \citep{testa2018} for this magnetar.

\subsubsection{CXOU J171405.7-381031}
This source was first discovered by \cite{aharonian2008} using the Chandra X-ray and High Energy Stereoscopic System (HESS) telescopes as a point source with a non-thermal spectrum coinciding with a radio shell and was identified as a magnetar candidate by \cite{halpern2010b}. With a followed up observation of Chandra, \cite{halpern2010} confirm the nature of the magnetar having the largest spin down of all magnetars and thus a characteristic age of  about \SI{1000}{yr}, making it a very young object.

\subsubsection{SGR 1627-41}
\cite{woods1999} discovered this magnetar with the Burst and Transient Source Experiment (BATSE) on the Compton Gamma Ray Observatory (CRGO) from several gamma ray bursts originating from the same sky position. It has been localised with high precision by \cite{wachter2004}. However, its spin period has only been discovered by \cite{esposito2009}, when the magnetar was showing an outburst in X-ray and was also associated with the SNR G337.0-0.1. These findings consolidated it as a member of the magnetar class.

\subsubsection{SGR 1806-20}
This SGR was first detected as a single gamma ray burst (GRB) in the KONUS experiment by \citep{mazets1981}. After detecting several more GRBs \cite{laros1987} proposed a common source, which they referred to as SGR 1806-20.
SGR 1806-20 has been one of the most active magnetars in terms of bursting activity with the peak being the giant flare emitted in 2004, which powers an expanding radio nebula \citep{hurley2005, gaensler2005b}. After this highly active phase, SGR 1806-20 has calmed down to a more quiescent state \citep{younes2017}.

\subsubsection{SGR 1833-0832}
This magnetar has initially been detected by \cite{barthelmy2010, gogus2010, gelbord2010} with observations of the Swift Burst Alert Telescope (BAT). It is one of the less studied sources with only one extensive study by \cite{esposito2011} showing a relatively high temperature and the typical magnetar timing behaviour. 

\subsubsection{SGR 1900+14}
\cite{kouveliotou1999} identified SGR 1900+14, which became active after a long period of quiescence, as a magnetar using the Rossi X-Ray Timing Explorer. This demonstrated that SGRs are indeed magnetars. As reported by \cite{hurley1999}, it is also one of the three magnetars that have been seen to emit giant flares, making it a great object for studying the fundamental physics of neutron stars.
It was initially found by \cite{mazets1979} as a source with a few gamma ray bursts before it is been identified as an SGR \citep{hurley1999b} with an associated Super Nova Remnant (SNR) \citep{vasisht1994}.

\subsubsection{SGR 1935+2154}
 It was originally discovered by \cite{stammatikos2014} and \cite{lien2014} using BAT through multiple soft gamma ray bursts and was associated with an SNR by \cite{gaensler2014}. Shortly after its discovery, an intermediate flare was detected \citep{kozlova2016}. This magnetar has been rather active with many glitches and bursts making it one of the most studied sources (for example \cite{younes2023}). Most prominantly, the discovery of an FRB like burst by \cite{bochenek2020} and \cite{chimesgr19352020} made SGR 1935+2154 the source that could potentially link FRBs to magnetars as progenitors. 
 %, which has been seen in later periods as well \cite{Hu2025}.

\subsubsection{Swift J1822.3-1606}
\cite{cummings2011} found this magnetar from repetitive soft gamma ray bursts from the same position on the sky using BAT and concluded that it is an SGR. The long term study of \cite{rea2012b} showed that this magnetar has one of the lowest magnetic field strengths in the magnetar sample (\SI{2.7e13}{G}), which is comparable to those of normal radio pulsars.
However, \cite{rea2012b} observed the typical magnetar X-ray outburst behaviour, which is the flux and spectral evolution, leading to the magnetar classification.

\subsubsection{Swift J1834.9-0846}
Swift J1834.9-0846 was first discovered by several gamma ray bursts from the same origin by \cite{delia2011} and \cite{guiriec2011} with BAT observations. Its nature as a magnetar was then confirmed by the detection of its spin period and derivative, which give the associated magnetic field strength of the magnetar during its 2011 outburst\citep{gogus2011, gogus2011b}.
%According to the fundamental plane of magnetars, this source should show radio emission, however the lack of the detection by \cite{tong2013} challenged this model for the prediction for radio emission.
Furthermore, it is associated with a SNR and was the first magnetar to be associated with a magnetar wind nebula \citep{younes2012}.

\subsection{Observational setup}
The observation were conducted in L and S1-band of MeerKAT, where the L-bands covers \SIrange{856}{1711}{MHz} and the S1-band covers \SIrange{1968}{2842}{MHz}.
In each band the data were recorded with a time resolution of \SI{32}{\micro s} and 1024 frequency channels with on coherent beam at the position of the source.
Of the 20 hours, 15 where observed in L-band and 5h in the S1-band.
As the radio loud magnetars undergo significant changes on time scales of weeks to months (for example \cite{bause2024} for XTE J1810-197), we decided to split our observations into seven epochs with a monthly cadence to observe the targets at as many different stages as possible.
The observations at L-band were taken on MJDs 60343, 60405, 60431, 60462 and 60550 while the S1-band observations where taken on MJDs 60372 and 60503.
Additionally, the magnetar 1E 1841-045 went into an X-ray outburst towards the end of the campaign, so we requested DDT time in S1 band for this source. These observations and results are described in \citep{younes2025}.

For each target we estimated the expected dispersion measure (DM) and scattering time $\tau$ in the L-band of MeerKAT using the NE2001 model. The DM estimate is used for the coherent dispersion at the time of observing.
While this is only an estimate, it potentially reduces in the intrachannel smearing significantly compared to using no coherent de-dispersion.
As the time in S1-band is limited, we target primarily those magnetars that have a high estimate of $\tau$ as the scattering will be reduce to approximately $(1.4/2.5)^4 \approx 0.1$ at S1-band.
Sources with $\tau$ larger than a few milliseconds at L-band, are targeted only in S1-band, while sources with negligible $\tau$ (less than \SI{0.2}{ms}) are targeted in L-band only.
In case of intermediate $\tau$ (around \SI{0.2}{ms}), they are targeted in both bands.
\Cref{tab:mag_properties} shows the estimated DM, $\tau$, observing bands and epochs for each target.

For the last two epochs, we observed in the commensal imaging and beam-forming mode. Hence, we observed polarisation (J1331+3030 | 3C286), flux (the closest to each source suggested by the MeerKAT OPT) and band (J1939-6342) calibrators for the imaging part in both epochs.
Additionally, a one minute scan of the test pulsar PSR J1602-5100 was added to each observation in order to have a clearly radio loud source for testing the pipelines.

\begin{table}
    \centering
        \caption{The estimated DM (in \si{pc \per cm^3}) used for the coherent de-dispersion, the estimated scattering at at MeerKAT L-band ($\tau$) in \si{ms}, the observing bands and at what epochs each target was observed. The DM of SGR 1935+2154 is taken from \citet{chimesgr19352020}.}
    \begin{tabular}{l|llll}
    \toprule
        Target & DM & $\tau$ & Bands & Epochs \\
         %& \si{km \per pc^3} & ms & & \\
        \midrule
        1E 1048.1-5937 & 648.2 & 0.2 & L, S1 & 1 - 7 \\
        1E 1841-045 & 1577.7 & 20.3 & L, S1 & 1 - 7, DDT\\
        1RXS J170849.0 & 1691.0 & 0.3 & L & 1,3,4,5,7\\
        3XMM J185246.6 & 1265.3 & 3.3 & L, S1 & 1-7\\
        CXOU J164710.2 & 1588.6 & 0.2 & L & 1, 3, 4, 5, 7\\
        CXOU J171405.7 & 1556.5 & 44.3 & S1 & 2, 6 \\
        SGR 1627-41 & 1692.5 & 65.3 & S1 & 2, 6\\
        SGR 1806-20 & 1548.2 & 21.7 & L, S1 & 1 - 7\\
        SGR 1833-0832 & 1827.7 & ? & L, S1 & 1 - 7\\
        SGR 1900+14 & 809.8 & 1.5 & L, S1 & 1 - 7\\
        SGR 1935+2154 & 332.7 & 0.3 & L, S1 & 1 - 7  \\
        Swift J1822.3-1606 & 1216.4 & 0.0004 & L & 1, 3, 4, 5, 7\\
        Swift J1834.9-0846 & 1630.5 & 0.2 & L, S1 & 1 - 7\\
        \bottomrule
    \end{tabular}
    \label{tab:mag_properties}
\end{table}

\section{Data reduction}\label{sec:dr}
Five epochs have been taken with only the beam forming data while for the last two observations we took commensal beam forming and imaging data. This section describes how both data sets have been processed in the search for radio emission.

\subsection{Beam forming}
The beam forming data of MeerKAT is in the form of \SI{8}{s} PSRFITS files, which are converted to total intensity filterbank files of the total observation duration for each source to ease the processing for the single pulse search and the folding.

\subsubsection{Single pulse search}
As all of the sources, with the exception of SGR 1935+2154, are so far radio quiet, the DM in the line of sight is unknown.
Our estimate only is a help for the coherent de-dispersion but the true value might be significantly off. Hence, we have to search a large DM range.
We decide to search a DM range of \SIrange{20}{10000}{pc \per cm^3}.
This avoids RFI, which has a DM of \SI{0}{pc \per cm^3} and covers a large potential DM range even beyond the estimates.
To ease the computational costs, a de-dispersion plan (DDplan) is created for each source that takes into account the smearing introduced as the DM moves away from the DM used for the coherent de-dispersion.
Thus, the data is down-sampled and searched with a more coarse DM grid when the smearing increases.
Additionally, the width of the single pulses is unknown.
We are searching in the range from the time resolution of the data to the period of the magnetar (known from X-ray observations) or \SI{10}{s} maximum pulse width.
To keep this computationally feasible, we split the search into four width ranges: from \SIrange{0.038}{2}{ms}, \SIrange{1}{100}{ms}, \SIrange{10}{1200}{ms} and \SIrange{0.16}{10}{s}, where each search is down-sampled in time by a multiple of 16 (1, 16, 256, 4096), respectively.
Hence, each of these width range requires its own DDplan.

The search is based on TransientX \citep{men2024}, while the creation of the DDplan is based on PRESTO's\citep{ransom2001}\footnote{\url{https://github.com/scottransom/presto}} \texttt{DDplan.py}.
To automate the search we developed \texttt{BLISS} (BLInd Single pulse Search), which creates the DDplans and starts the searches for each source.
The search makes use of the inbuilt RFI mitigation techniques from TransientX (the skewness-kurtosis filter) and  we additionally apply the \texttt{zdot} filtering, which is an improved zeroDM removal filtering and the kadaneF filtering \citep{men2023}.
In the search, we use 6 as a threshold for the signal to noise ratio (S/N). This value is based on the false alarm rate for the typical observation length and the widths searched, so that the number of expected noise candidates is of the order of 10 pulses per observation.
Moreover, \texttt{BLISS} also calculates the modulation index and spectral kurtosis for each candidate for filtering out RFI as described in \cite{bause2024}.
After the search, those candidates below the threshold for the modulation index are visually inspected for a detection of an astrophysical signal.
The threshold can be estimated based on the S/N threshold and the frequency resolution as described in \cite{spitler2012} as $\approx 1$. Taking into account that the signals might not be completely broadband, as it can be the case for FRBs and bursts from SGR1935+2154, we apply a slightly higher modulation index threshold, that is we accept more candidates, of 1.1 (L-band) and 1.5 (S-band). Each of the remaining candidates is then inspected visually.

\subsubsection{Folded profiles}
In addition to the search for single pulses, we are searching for the pulsed emission as it is typically done for pulsars, for which the period, period derivative and DM are required.
While the period and the period derivative are known from X-ray observations, the DM is, as for the single pulse search, unknown.
We cover the same DDplan as for the single pulses (range: \SIrange{20}{10000}{pc \per cm^3}) and fold our data around the known periods in using with a Fast Fourier Transform (FFT) approach (peasoup\footnote{\url{https://github.com/ewanbarr/peasoup}}) and a Fast Folding Algorithm (FFA) approach (\texttt{riptide} \cite{morello2020}).
Prior to folding the observations, the filterbank files are being RFI cleaned by \texttt{filtool}\citep{men2023} using the same RFI mitigation as in the TransientX single pulse search.
%In the folding process, masks are used that are created using \texttt{rfifind} with the default parameters, while setting the time range to \SI{1}{s} that are generated for each filterbank file. To avoid artifacts from the imaging observations, the first \SI{10}{\%} of each observation with commensal imaging are neglected.
%For each source and observation, two folds are done: one using a coarse search, one using a fine search.
%The resulting folded profiles are inspected visually for a detection.
The resulting candidates are inspected for a detection close to the expected period of the respective magnetar.

\subsection{Imaging}\label{sec:imaging}

\subsubsection{Total intensity}
\begin{figure*}
    \centering
    \includegraphics[width=0.45\textwidth]{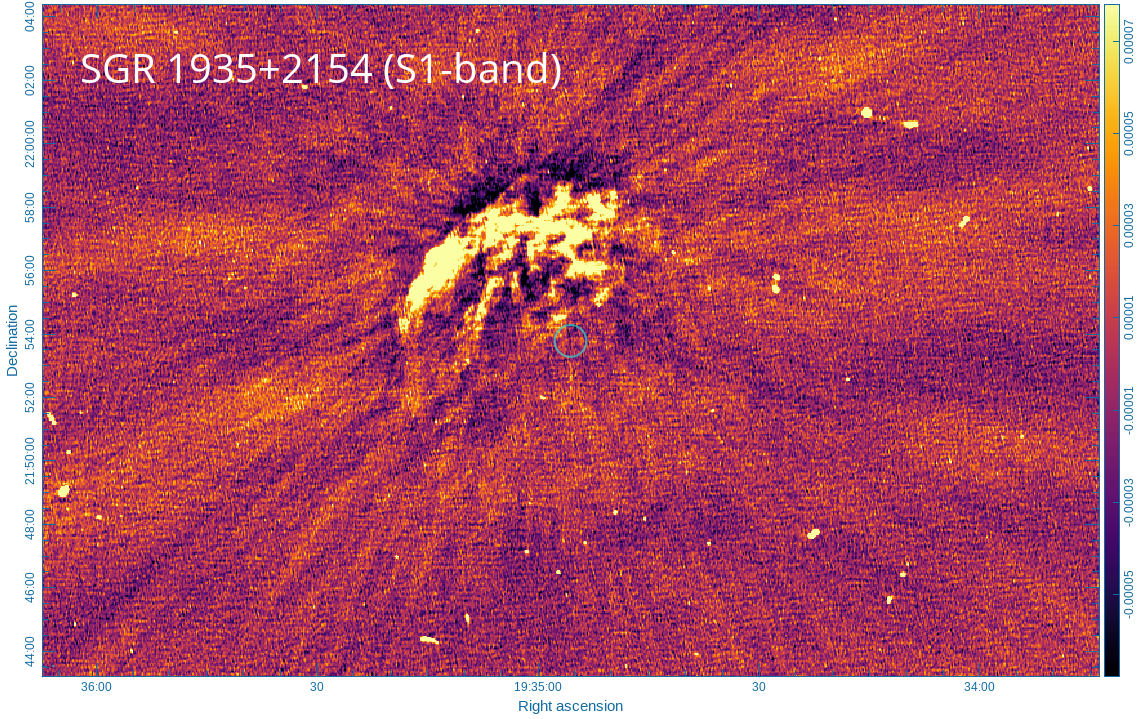}
    \includegraphics[width=0.45\textwidth]{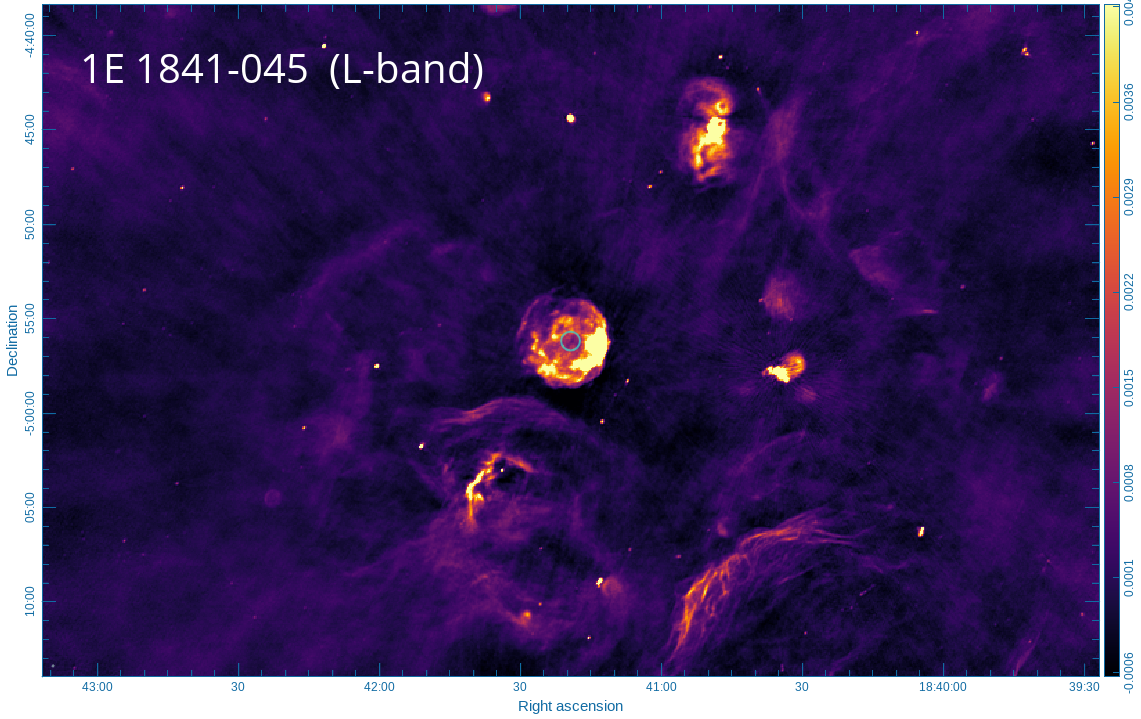}
    
    \caption{Total intensity images of two two magnetars with a SNR association. Left: SGR 1935+2154 at the S1-band, right: 1E1841-054 at L-band.}
    \label{fig:meerkat_images}
\end{figure*}

The imaging for the total intensity images is done using the oxkat pipeline \citep{oxkat}.
This pipeline provides among others routines for the reference calibration (1GC), the flagging of bad baselines and the direction dependents self calibration (2GC). We follow the 1GC and flagging procedures with the default parameters.
For the 2GC calibration, we do an iterative imaging procedure going down from \SI{64}{s} integration time to \SI{32}{s}, \SI{16}{s} and finally \SI{8}{s} to account for time-dependent phase changes.
In each step, we create a mask with a threshold chosen by visual inspection to mask only the astrophysical sources and imaging with cleaning using the default parameters of oxkat despite the number of iterations (80,000) and enabling multi scale cleaning (scales 0, 3 and 9).
The final images are then primary beam corrected using oxkat and are searched for point sources at the position of the magnetar. \Cref{fig:meerkat_images} shows two examples for the resulting images while the remaining images are presented in \Cref{fig:mkt_lband_all}.

In addition to the magnetars presented above, we make use of the fact that the ultra long period object ASKAP J1935+2148 and the magnetar SGR1808-20 have serendipitously observed during our SGR 1935+2154 and SGR 1806-20 observations respectively.
Hence, we can apply the imagining search methods to these as well.
During the imaging, we noted that the sources SGR 1627-41 and CXOU J171405.7-381031 haven been observed with an offset larger than \SI{1}{\degree}.
Hence, the data for the imaging epoch is not used. This includes also the beamforming data for these two sources.

\subsubsection{Polarisation images}
The two imaging epochs were also observed with a polarisation calibrator. Thus, we can create Stokes V images. This is useful as diffuse emission is often bright in Stokes I but (almost) invisible in Stokes V, while the few objects that are emitting circular polarisation, which includes magnetars and pulsars, will stand out from their environment.
Thus we can create the Stokes V images, we make use of the pipeline developed for the Max Planck Institute for Radio Astronomy MeerKAT Galactic Plane Survey \citep{mmgps2023} (MMGPS)\footnote{\url{https://www.mpifr-bonn.mpg.de/mmgps}}, since oxkat does not yet support this step. 
%As   Moreover, circular polarization imaging provides an independent search channel for pulsar candidates and, by extension, magnetars, which likewise appear as point sources.
%Although we did not detect any magnetars in our Stokes V images, we did recover a pulsar, thereby validating our polarization calibration and imaging methodology. 

\subsubsection{Imaging light curves}
\begin{figure}
    \centering
    \includegraphics[width=\linewidth]{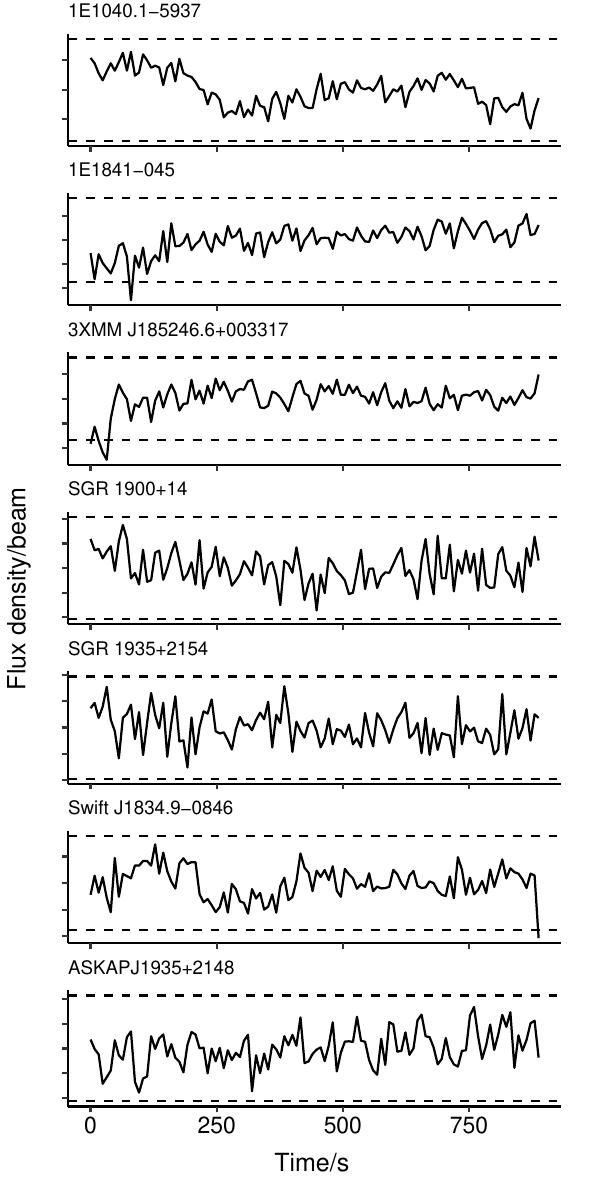}
    \caption{Light curves (solid lines) and +-3 sigma limits (dashed lines) for each source for the S1-band observation for the magnetars as well as the ULP ASKAP J1935+2148, which is in the beam of the SGR 1935+2154 observation. The time resolution is \SI{8}{s}.}
    \label{fig:lightcurves_sband}
\end{figure}

%The search for pulses of the order of seconds are not very sensitive in time domain as among other the dispersive delay is much shorter than the duration of a pulse, making the detection and classification challenging.
Given that the images where taken in \SI{8}{s} intervals, we can create a light curve of the the position of the source and thus search for emission that is similar to how ultra long period transients are searched.
This also helps against RFI signals, which originate from Earth.
For creating these so-called snapshot images, we use the fully calibrated total intensity images and create images without cleaning but subtracting the model to focus on the transient emission only.
Given the length of the observation, we get 112 images per source, which correspond to 112 time bins.
For each image, we average the area with a \SI{5}{\arcsecond} radius around the source position. 
This area is somewhat smaller than the typical beam for a point source as we focus on the bright centre of the source's position.
The results are the light curves for each magnetar and ASKAP J1935+2148 displayed in \cref{fig:lightcurves_sband} and \cref{fig:lightcurves_lband}, which we inspect visually for any signal higher than three times the standard deviation after subtraction of the mean of the light curve.

\section{Results}\label{sec:results}
\subsection{Beam forming}
\begin{figure}
    \centering
    \includegraphics[width=0.85\linewidth]{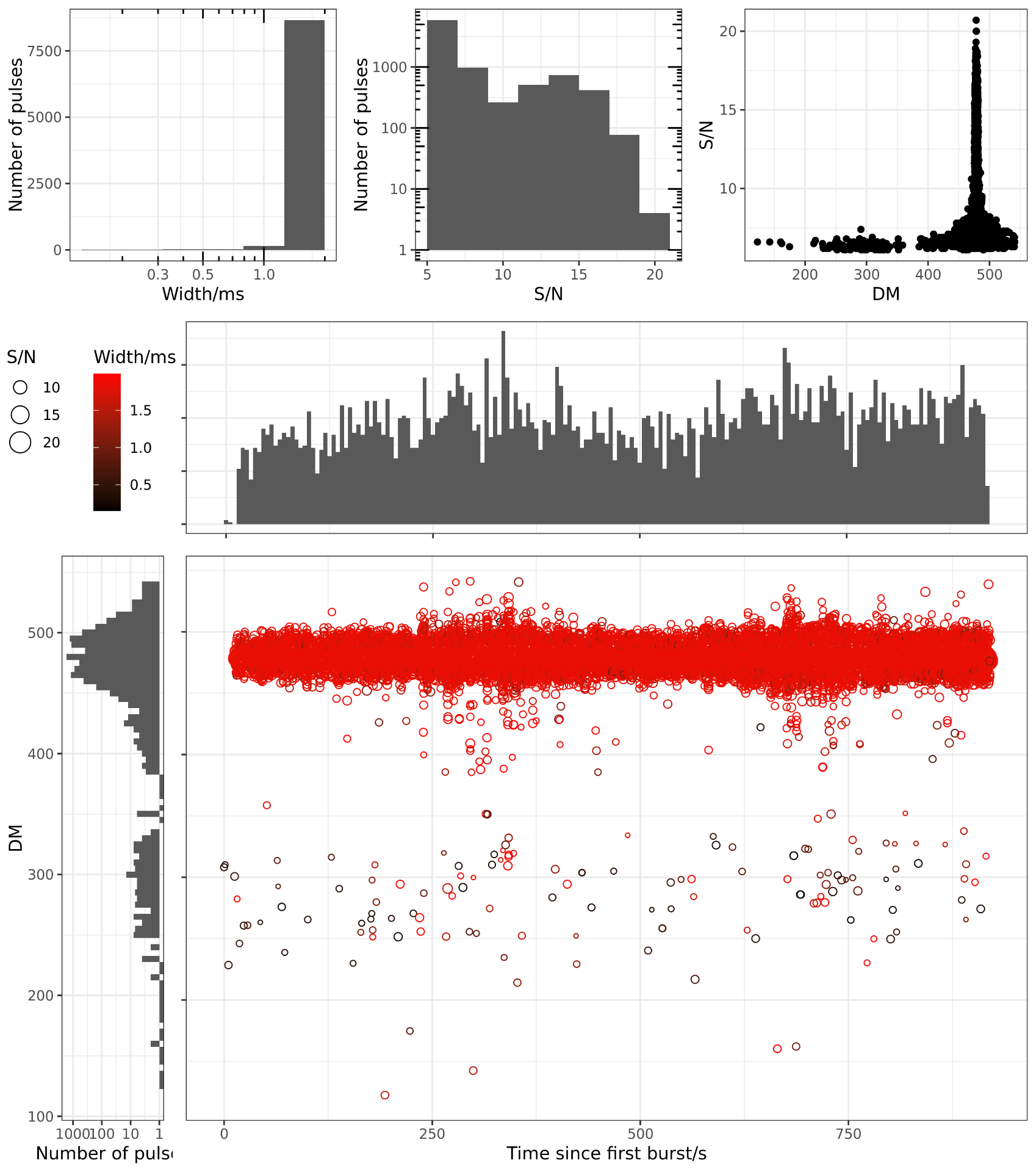}
    \caption{Over view of the single pulse detections of the pulsar in the Beam of CXOU1647 in an L-band observation. The top row of panels shows (from left to right) a histogram of the detected single pulse widths and S/N respectively as well as S/N vs. DM. The bottom part shows a DM vs. time scatter plot of all candidates, where each axis as a histogram added and the S/N of the candidates are given by the circle size while the width is given by the colour of the circles.}
    \label{fig:cxoupulses}
\end{figure}

In the two beamforming searches no single pulse or folded profile that can be related to the magnetars has been found above the S/N threshold of 6.
Nevertheless, our observations showed several notable re-detections of pulsars in the proximity of our targeted magnetars both in single pulses as well as folding searches.
The most notable is the detection of the pulsar B1641-45 in the beam of CXOU J164710.2-455216, which has been seen when folding with the period of the magnetar itself but also many single pulses were discovered (\cref{fig:cxoupulses}). Given the DM of \SI{478}{pc \per cm^3} and the periodicity of \SI{455}{ms}, these pulses can be easily related to PSR B1641-45. Its angular distance is \SI{0.42}{\degree} from the from CXOU J164710.2-455216.
Additionally, it clearly demonstrates that our pipeline \texttt{BLISS} is able to blindly find single pulses of a source in the beam.
On the other hand, this clearly limits the detectability of single pulses from the magenetar itself as the period of the pulsar is about \SI{4}{\%} of the period of the magnetar.
Additionally, as both sources are in the proximity of the Westerlund I cluster, their DM is potentially also quite similar.
This makes distinguishing between a single pulse from the magnetar and the pulsar challenging.

We additionally found single pulses from a pulsar when inspecting the data from the magnetar SGR 1627-41. As these were three pulses separated by about \SI{440}{ms} with a DM of \SI{470}{pc \per cm^3}, they can be related to the pulsar PSR B1630-44, which happens to be in the beam due to the spurious \SI{1}{\degree} offset in this observation.

For our main sources, the magnetars, we can thus only give upper limits for the radio fluence of single pulses and the mean radio flux density of the folded profiles.
We estimate the upper limit mean flux density $S_{mean}$ and the upper limit single pulse fluence $F_{SP}$ using the respective version of the radio meter equation:
\begin{align*}
S_{mean} &= \frac{(S/N) Z T_{sys}}{G \sqrt{n_{pol} t_{obs} B}} \sqrt{\frac{X}{1-X}} \\
F_{SP} &= \frac{(S/N) T_{sys} \sqrt{w}}{G \sqrt{n_{pol} B}},
\end{align*}
where $(S/N) = 7$ is the minimal signal to noise ratio of the folded profile or the single pulse respectively, $T_{sys}$ is the system temperature, $G$ is the gain, $n_{pol} = 2$ is the number of polarisations recorded, $t_{obs}$ is the duration of the observation, $B$ is the bandwidth of the receiver, $X$ is the duty cycle, $Z$ is the correction factor for the red noise contribution and $w$ is the single pulse width.
We adopt the same values as used in \cite{younes2025} for the telescope dependent properties.
That is $G$=\SI{2.65}{K/Jy} and $T_{sys}$ = \SI{26}{K} for the L-band observations and an SEFD=$T_{sys}/G$ for 56 antennas of \SI{8.6}{Jy} for the S1-band observations.
To estimate $Z$, we compare the noise levels (RMS) of the residual FFT on the millisecond regime to the RMS in the regime of the period of the slowest magnetar (\SI{12}{s}).
This gives a factor of about 4 and thus, $Z \approx$ 4.

For the regular L-band and S1-band observations we estimate $S_{mean} = \SI{68}{\micro Jy}$, $F_{SP} = \SI{52}{\milli Jy~ \milli \second}$ and $S_{mean} = \SI{52}{\micro Jy}$, $F_{SP} = \SI{39}{\milli Jy~ \milli \second}$ respectively.
The last two observations with commensal imaging (one at each band) were slightly longer and thus the upper limits for $S_{mean}$ are \SI{60}{\micro \Jansky} (L-band) and \SI{43.6}{\micro \Jansky} (S1-band).
Using the distance estimates of the magnetars, we can estimate the corresponding luminosity and single pulse energy as upper limits. These values are listed in \Cref{tab:mag_upperlimits}.
\begin{table*}
\centering
    \caption{Overview of the upper limits from both time and imaging domain. For each target the assumed distance (from the magnetar catalogue and \citet{caleb2024}), the spectral luminosity $L_{mean}$ for the folded profile, the spectral single pulse fluence $F_{SP}$, the flux density limits from the total intensity $F_{Im}$ and the corresponding spectral luminosity $L_{Im}$ as well as the flux density upper limits from the light curves $F_{ULP}$ and the corresponding spectral luminosity $L_{ULP}$ are given for both the L and S1-band observations. Due to the offset of the S1-band imaging observation, SGR 1627-41 and CXOU J171405.7 do not have any imaging upper limits and the time domain ones are only fore epoch 2.}
    \begin{tabular}{l|l|llllll|llllll}
\toprule
         & & \multicolumn{6}{c}{L-band} & \multicolumn{6}{c}{S1-band}\\
        Target & D & $L_{mean}$ & $F_{SP}$ & $F_{Im}$ & $L_{Im}$ & $F_{ULP}$ & $L_{ULP}$ & $L_{mean}$ & $F_{SP}$ & $F_{Im}$ & $L_{Im}$ & $F_{ULP}$ & $L_{ULP}$\\
        Prefactor  & -- & $10^{13}$ & $10^{13}$  & $10^{-3}$  & $10^{13}$  & $10^{-3}$  & $10^{13}$  & $10^{13}$  & $10^{13}$  & $10^{-3}$  & $10^{13}$  & $10^{-3}$  & $10^{13}$ \\
       Units  & kpc & \fracsi{\erg\per\hertz} & \fracsi{\erg\per\hertz\per\second} & \si{\Jansky} & \fracsi{\erg\per\hertz\per\second} & \si{\Jansky\second} & \fracsi{\erg\per\hertz} & \fracsi{\erg\per\hertz} & \fracsi{\erg\per\hertz\per\second} & \si{\Jansky} & \fracsi{\erg\per\hertz\per\second} & \si{\Jansky\second} & \fracsi{\erg\per\hertz}\\

\midrule
%1E 1048.1-5937       & 9.0 & 5.29 & 1.73 & 0.00 & 3.79 & 0.00 & 254.47 & 3.97 & 1.32 & 0.00 & 1.36 & 0.00 & 91.61 \\
%1E 1841-045          & 8.5 & 4.72 & 1.54 & 0.01 & 59.58 & 0.00 & 54.48 & 3.54 & 1.18 & 0.00 & 0.37 & 0.00 & 36.32 \\
%1RXS J170849.0       & 3.8 & 0.94 & 0.31 & 0.00 & 0.49 & 0.00 & 14.52 & --   & --   & --   & --   & --   & --   \\
%3XMM J185246.6       & 7.1 & 3.29 & 1.08 & 20.00 & 20.00 & 20.00 & 20.00 & 2.47 & 0.82 & 0.00 & 0.44 & 0.00 & 19.00 \\
%CXOU J164710.2       & 3.9 & 0.99 & 0.32 & 0.00 & 0.56 & 0.00 & 11.47 & --   & --   & --   & --   & --   & --   \\
%CXOU J171405.7       & 13.2 & --   & --   & --   & --   & --   & --   & 8.54 & 2.85 & --   & --   & --   & --   \\
%SGR 1627-41          & 11.0 & --   & --   & --   & --   & --   & --   & 5.93 & 1.98 & --   & --   & --   & --   \\
%SGR 1806-20          & 8.7 & 4.95 & 1.62 & 0.01 & 69.91 & 0.00 & 95.11 & 3.71 & 1.24 & --   & --   & --   & --   \\
%SGR 1833-0832        & 10.0 & 6.53 & 2.14 & 0.00 & 0.90 & 0.00 & 125.66 & 4.90 & 1.63 & --   & --   & --   & --   \\
%SGR 1900+14          & 12.5 & 10.21 & 3.34 & 0.00 & 5.15 & 0.00 & 98.17 & 7.66 & 2.55 & 0.00 & 0.32 & 0.00 & 39.27 \\
%SGR 1935+2154        & 9.0 & 5.29 & 1.73 & 0.00 & 4.27 & 0.00 & 40.72 & 3.97 & 1.32 & 0.00 & 0.18 & 0.00 & 20.36 \\
%Swift J1822.3-1606   & 1.6 & 0.17 & 0.05 & 0.00 & 0.76 & 0.00 & 4.18 & --   & --   & --   & --   & --   & --   \\
%Swift J1834.9-0846   & 4.2 & 1.15 & 0.38 & 0.00 & 2.04 & 0.00 & 8.87 & 0.86 & 0.29 & 0.00 & 0.22 & 0.00 & 11.08 \\
1E 1048.1-5937     & 9.0  & 5.29 & 6.92 & 0.60 & 60.56  & 2.50 & 254.5 & 3.97 & 5.29 & 1.34 & 136.1 & 0.90 & 91.61 \\
1E 1841-045        & 8.5  & 4.72 & 6.17 & 6.30 & 572 & 0.60 & 54.48  & 3.54 & 4.72 & 0.41 & 37.24  & 0.40 & 36.32 \\
1RXS J170849.0     & 3.8  & 0.94 & 1.23 & 0.43 & 7.76   & 0.80 & 14.52  & --   & --   & --   & --     & --   & --    \\
3XMM J185246.6     & 7.1  & 3.29 & 4.31 & 2.45 & 155.2 & 0.60 & 38.01  & 2.47 & 3.29 & 0.70 & 44.25  & 0.30 & 19.00 \\
CXOU J164710.2     & 3.9  & 0.99 & 1.30 & 0.47 & 9.00   & 0.60 & 11.47  & --   & --   & --   & --     & --   & --    \\
CXOU J171405.7     & 13.2 & --   & --   & --   & --     & --   & --     & 8.54 & 11.4 & --   & --     & --   & --    \\
SGR 1627-41        & 11.0 & --   & --   & --   & --     & --   & --     & 5.93 & 7.91 & --   & --     & --   & --    \\
SGR 1806-20        & 8.7  & 4.95 & 6.47 & 11.8 & 1119 & 1.00 & 95.11  & 3.71 & 4.95 & --   & --     & --   & --    \\
SGR 1833-0832      & 10.0 & 6.53 & 8.55 & 0.11 & 14.43  & 1.00 & 125.7 & 4.90 & 6.53 & --   & --     & --   & --    \\
SGR 1900+14        & 12.5 & 10.2 & 13.35 & 0.42 & 82.47  & 0.50 & 98.17  & 7.66 & 10.21 & 0.16 & 32.16  & 0.20 & 39.27 \\
SGR 1935+2154      & 9.0  & 5.29 & 6.92 & 0.67 & 68.33  & 0.40 & 40.72  & 3.97 & 5.29 & 0.17 & 17.60  & 0.20 & 20.36 \\
Swift J1822.3-1606 & 1.6  & 0.17 & 0.22 & 3.79 & 12.21  & 1.30 & 4.18   & --   & --   & --   & --     & --   & --    \\
Swift J1834.9-0846 & 4.2  & 1.15 & 1.51 & 1.47 & 32.59  & 0.40 & 8.87   & 0.86 & 1.15 & 1.01 & 1.40   & 0.50 & 11.08 \\
ASKAP J1935+2148   &5	  &-	 &-	    &0.34  & 10.64  &0.4   & 12.57  & -    & -    &	0.17 & 0.33   &	0.20 & 6.28\\
SGR1808-20         &	10 &-	 &-	    &0.77  & 96.76  &0.8   & 100.5 &	-  &-     & -    &	-     & -    &-\\

\bottomrule
    \end{tabular}

    \label{tab:mag_upperlimits}
\end{table*}
%\begin{itemize}
%    \item Upper limits
%    \item Notable side detections, e.g. pulsar pulses etc.
%\end{itemize}

\subsection{Imaging}
\Cref{fig:meerkat_images} shows two examples of the resulting total intensity images: S1-band for SGR 1935+2154 and L-band for 1E1841-045.
For each of the two sources, the associated supernova remnant is clearly visible. 
For the SGR 1935+2154 images, we also marked the position of the ultra long period transient discovered by \citet{caleb2024}.
%For both sources the associated super nova remnant is visible however also not emission at the position of the source is visible.
%In the images of every magnetar (both at Stokes I and Stokes V), we search for point sources at the position of the magnetars by visually inspecting the images.
For both sources, the associated supernova remnant is visible.
Considering the magnetars as point sources, we searched for them as their respective positions by inspecting the Stokes I and Stokes V images. However, we did not find any radio point-source association for any of the magnetars considered
%However, we do not find a radio point source that we can associate with the magnetar in our images. 
In the case of SGR 1806-20, we find some emission at the position of the source.
After further inspection, it appears most likely to be diffuse emission of the surrounding ISM due to the non point source like appearance of the emission. Moreover, there is no emission at this position in Stokes V.
Hence, we derive upper limits from the imaging domain on the persistent radio emission of each magnetar.
To do so, we estimate the RMS of the magnetar in a circle with a radius of \SI{30}{\arcsecond} around the magnetar's position. To be detectable, we require a source to be at least seven times the RMS, which is a threshold inferred by other point sources in the images.
The corresponding values for the total intensity images are listed in \cref{tab:mag_upperlimits}, where the luminosity using the distance is also estimated. For the sources, which are located in an area of diffuse emission (such as SGR 1806-20), the upper limits is higher as a consequence.
As the diffuse emission is (almost) completely gone in the Stokes V images, here the upper limit for non-detections is the same among the different sources and about \SI{2e-4}{Jy/beam} in L-band and \SI{2.1e-4}{Jy/beam} for the S1-band observation respectively.

For the light curves obtained from the images displayed in \cref{fig:lightcurves_lband} and \cref{fig:lightcurves_sband}, we do not find any signal above our detection threshold in the imaging time series, despite an outlier for SGR 1935+2154.
The inspection of the corresponding images and beamforming candidates around the outlier reveals no no signal of interest.
Applying the somewhat more conservative threshold for a detection of seven times the RMS, we estimate an upper limit for each light curve for the flux density and luminosity respectively.
Both are listed in \Cref{tab:mag_upperlimits}.

Despite the non-detection of radio emission from our main targets (the magnetars) we can clearly see that the environment of these sources is very complex, which explains the rather high DM estimates for most of the sources.
We find several pulsars in the imaging domain, such as PSR J1841-0500 in the S1-band observation, which is an intermittent pulsar with a substantial amount of circular polarization \citep{camilo2012}.
As some of the pulsars are visible in the circular polarisation images, we can confirm the quality of our images and that the non-detections of the magnetars in the Stokes V images are indeed astrophysical.

Additionally, the images give a view the environment of the magnetars themself and their associated structures such as SNRs.
All images of the sources are presented in \Cref{fig:meerkat_images} and \Cref{fig:mkt_sband_all}.
Clearly, the SNRs of SGR 1935+2154 and 1E1841-045 can be identified in their images at both L- and S-band.
Even the other magnetars are located in complex environments, which given their location close to the galactic plane, is not surprising.
%\begin{figure*}
%    \centering
%    \label{fig:meerkat_images}
%    \caption{Examples of images from MeerKAT}
%\end{figure*}

\section{Discussion}\label{sec:discussion}

The non-detection of any radio signal from the targeted magnetars can be due to 1.) that they are indeed radio quiet during our observations or 2.) that there was emission that we could not detect.
Reasons for the latter include a weak source that we were not sensitive to, an unfortunate beaming angle, effects of the interstellar medium, such as scattering or diffuse emission (for imaging only), which diminish the signal from the source or a combination of all of them.

While the general emission mechanism of magnetars is unclear, there are some arguments one can try to make to rule out radio emission from magnetars based on their physical properties.
\citet{szary2015} extend the partial screen gap model from radio pulsars into the magnetar regime making the radio emission rotationally powered and thus can make predictions on whether a magnetar emits radio or not based on the temperature in its quiescent state and the period and period derivative.
Based on the arguments made by \citet{szary2015} the following magnetars should not emit radio emission: 1E 1048.1-5937, 1E 1841-045, 1RXS J170849.0, CXOU J164710.2, SGR 1806-20, SGR 1900+14 and Swift J1822.3-1606 either because of a too high surface temperature or too strong magnetic fields.

In addition to the traditional emission mechanism powered by rotational energy, emission mechanisms powered by the strong magnetic fields have been proposed.
\citet{wang2019} argue that the emission of the radio loud magnetars XTE J1810-197 and PSR J1622-4950 is powered by oscillations in the magnetosphere, which are induced by a quake of the crust of the magnetar. Additionally, \cite{wang2024} argue that the radio emission seen in SGR 1935+2154 during the detection of several weak radio pulses is powered by the untwisting magnetic field in the outer magnetosphere. 
\citet{cooper2024} extend this model to ultra long period transients arguing that they are magnetically powered but also noting that the normal magnetars, that is the magnetars considered in this work, are too young and might not have built up enough twist in the magnetosphere to sustain the magnetic radio emission.
As the death lines and the fundamental plane of magnetars \citet{rea2012} are based on rotational powered radio emission, these cannot make predictions on the radio emission on magnetars that are powered on the magnetic field.

While the process that initiated the radio emission differs between rotationally powered and magnetically powered models, the opening angle of the open field line region (or its equivalent in case of magnetically powered radio emission) defines the beaming fraction (as discussed by \citet{beloborodov2009}), that is the region of the sky illuminated by the beam.
The beaming fraction is thus another limiting factor in the detectability of radio emission.
While for ordinary radio pulsars this is commonly estimated from the open field line polar cap geometry and empirical fits (for example\citet{tauris1998}), magnetars do not generally follow the narrow polar cap picture.
As shown by for example \citet{philippov2022}, the beam widths of magnetars are about an order of magnitude larger than those of radio pulsars, which indicates that also their beaming fractions are much larger (these would correspond to a beaming fraction of ~\SI{4}{\%} if the pulsar-based polar cap model were applied to these periods.).
Thus, the non-detection of radio emission from all 13 magnetars is therefore unlikely to be due solely to geometrical effects.

%In the case of the magnetically powered models, the effective open field line is further constrained by the twisting of the magnetosphere, which changes over times as discussed by \citet{wang2019} and thus adds a time dependency to the beaming fraction. Therefore, the above calculation requires information about the current magnetic field configuration, which is not present.

Our understanding of magnetar radio emission is heavily biased due to the small sample of known radio loud magnetars.
Several assumptions such as the radio follows an X-ray event relation might not be true as seen by the re-brightening of XTE J1810-197 without notable X-ray activity \citep{caleb2022,bause2024} or 1E 1547.0-5408, which turned radio off after an X-ray outburst \citet{lower2023}.
%Additionally, some radio loud magnetars like SGR J1745-2900 in the Galactic centre are located in such special environments so that other effects are playing a relevant role, potentially also for capability of emitting in the radio regime,
Additionally, some radio loud magnetars are emitting in radio only very sporadically like SGR 1935+2154.
Thus defining a a "regular" radio loud magnetar is challenging as so far it seems like a class of many special cases.
%Thus, , which not even for arguably the simpler case of radio pulsar emission has been found is a challenging task and predicting radio emission from magnetars is yet an uncertain forecast.

Additional challenges in detecting the radio emission from magnetars are there distances, making them faint, and their environment, which influences the emitted radio emission.
The estimated DMs and $\tau$ for the magnetars are very high compared to the typical galactic radio pulsar, which is challenging in the search for radio emission in the time domain, especially at lower frequencies due to strongly scattered radio emission.
This is due to their tendency of being in more complex structures, where a lot of material is in the lines of side to us as it can be seen in the images in this work.
$\tau$ is highly frequency dependent and thus changing to higher frequency reduced the effect of scattering significantly.
While it is generally assumed that magnetars have flat spectra in contrast to radio pulsars, this finding originates from the small set of magnetars, where enough radio emission has been observed.
Hence, moving to higher frequencies to avoid scattering might still cause a reduced brightness if the spectrum if is not flat or has a turn over as seen for XTE J1810-197 \citep{maan2022}.
In this particular case, the turn over is thought to originate from free-free absorption, which limits the detectable radio flux. Given that a large number of magnetars is located in dense regions of the Milky Way and free-free absorption can happen, it could impact their detectability as well.
The imaging domain on the other hand does not suffer from scattering and is still able to detect pulsed emission that has been scattered so strongly that the profile or single pulses are not detectable anymore in beamformed searches.

\begin{figure}
    \centering
    \includegraphics[width=\linewidth]{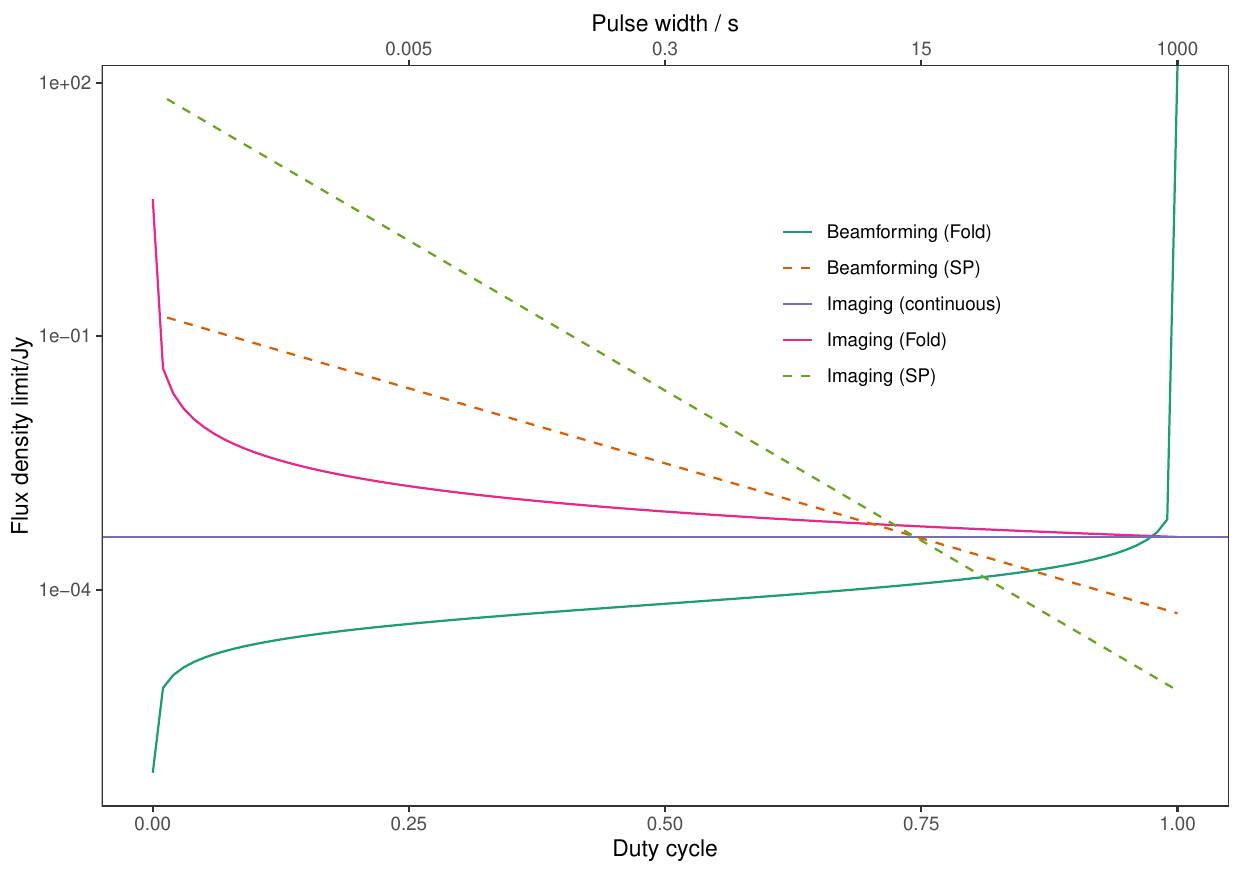}
    \caption{Overview of the different flux density limits obtained for the sources. Here, we are displaying the limits for 1RXS J170849 at L-band but the other sources and bands are analogue. The solid lines correspond to the limits on the pulsed (pulsar-like) emission from imaging ($F_{IM}/X$, where $X$ is the duty cycle) and beamforming ($S_{mean}$) as a function of duty cycle (linear scale). The dashed lines show the limits for the single pulses (log scale): $F_{SP}/w$ for beamforming and $F_{ULP}\frac{\SI{8}{s}}{w}$ for imaging ($w$: pulse width).}
    \label{fig:limit_lines}
\end{figure}

We have searched for radio emission using four different search techniques. In \cref{fig:limit_lines}, we are presenting the limits on the (peak) flux density as a function of duty cycle (pulsar-like emission) from both the imaging and the beam forming perspective, the limit on persistent radio emission (equal to a duty cycle of 100\%) and for single pulses using the beam forming search, as well as the imaging time series, assuming no scattering of the emission.

Clearly, the time domain is more sensitive to short pulses and small duty cycles as these will be averaged out by the low sampling rate of the imaging domain.
Only on the scale of \SI{10}{s} and > 90\% duty cycle, the imaging domain provides stronger limits for this source (1RXS J17084).
However, for longer pulses and duty cycles other effects, such as strong red noise or base line variations, that are not considered in our estimates start to play a role.
Thus the turnover, where the imaging domain, which is also not affected by scattering, is taking over is likely to occur earlier than it is demonstrated here. 
Moreover, an increased sampling rate of the snapshot imaging would increase the sensitivity to single pulses from the imaging domain.
%For sporadic, short pulses that are much shorter that the imaging rate, the image domain is however less sensitive then a folded profile emission.
Thus, both domains combined are key in the search for radio emission from these objects, as both provide limits on a different part of the parameter space of possible radio emission.
The ongoing and upcoming all-sky surveys such as the ASKAP Variables and Slow Transients survey (VAST \citet{tara2013}), BURSTT \citep{lin2022}, the SKA, CHORD \citep{vanderlinde2019} or DSA 2000 are ideal instruments that can deliver a very high cadence of observations of a large sky fraction that also includes positions of many radio quiet magnetars. We encourage these instruments to also monitor for any radio activity from magnetars using both the imaging and time domain.

\section{Conclusion}\label{sec:summary}

In this work we present the results of an eight-month-long monitoring campaign of twelve radio quiet magnetars and SGR 1935+2154 using the MeerKAT interferometer radio telescope.
While focused on the the time domain, we also make use of MeerKAT's interferometric nature and the powerful data taking backend for a search for radio emission in the image domain. Thus, we can search for radio emission from four different perspectives:
\begin{enumerate}
    \item The FRB perspective by blindy searching for (faint) single pulses of widths from the time resolution to the period of the magnetar over a DM range from \SIrange{20}{10000}{pc \per cm^3}. The non-detections give upper limits of \SI{52}{mJyms} (L-band) and \SI{39}{mJyms} (S1-band) for a \SI{10}{ms} burst.
    \item The radio pulsar perspective by folding the observations to search for pulsed emission using both an FFT and an FFA approach. For the L-band observations we can report upper limits of \SI{68}{\micro Jy}, while for the S1-band we can report upper limits of \SI{56}{\micro Jy} for each source during out observational campaign.
    \item The persistent radio emission perspective by searching for point sources at the positions of the magnetars in the epochs, where the full calibrator scheme for the imaging domain was followed. Our upper limits depend on the environment of the source and are of the order of $\approx$\SI{0.5}{mJy/beam} for most sources.
    \item The ultra long period transient or long duty cycle perspective, that is searching for single pulses in the image domain by using \SI{8}{s} snapshot images at the position of the magnetars and generating a light curve for each source. The upper limits are about $\lessapprox$ \SI{0.1}{mJy/beam}.
\end{enumerate}
In addition to the 13 magnetars specifically targeted, we make use of the field of view of MeerKAT in the imaging domain to report also upper limits for the ultra long period transient ASKAP J1935+2148 and the magnetar SGR 1808-20 and re-detect several radio pulsars.

We find our results partially in agreement with the models trying to predict radio emission from magnetars but argue that the foundation of magnetars used in magnetar emission models might not be sufficient due to several magnetars being special cases.
Hence, we highly encourage follow up and reports of non detection of radio quiet magnetars with high cadence independent of X-ray activity of the magnetar. Ideally searching in as many for the four perspectives as possible to gain a broader view on the magnetars potential of emitting in the radio regime and thus helping to understand their emission mechanism as well as their connected phenomenons such as FRBs.

\begin{acknowledgements}
L.G.S. is a Lise Meitner Group Leader, and together with M.L.B. acknowledge support from the Max Planck Society. The MeerKAT telescope is operated by the South African Radio Astronomy Observatory, which is a facility of the National Research Foundation, an agency of the Department of Science and Innovation. This work has made use of the ‘MPIfR S-band receiver system’ designed, constructed, and maintained by funding of the MPI für Radioastronomy and the Max Planck Society. Observations used PTUSE for data acquisition, storage, and analysis which was partly funded by the Max-Planck-Institut für Radioastronomie (MPIfR). We acknowledge the MMGPS pipeline\footnote{https://www.mpifr-bonn.mpg.de/mmgps} development team for helping in part with the data analyses.
\end{acknowledgements}

   \bibliographystyle{aa} % style aa.bst
   \bibliography{refs.bib} % your references Yourfile.bib

\begin{appendix}
 \section{Data availability}
The final images presented in this work are available in the CDS. The raw measurement sets taken from MeerKAT as well as the raw time domain data are available from the MeerKAT data archive. The intermediate filterbanks and light curves are stored at the archive of the Max-Planck-Institute for radio astronomy and can be shared up on reasonable request.

\section{MeerKAT images and light curves of all sources}
\begin{figure}
    \centering
    \includegraphics[width=0.45\textwidth]{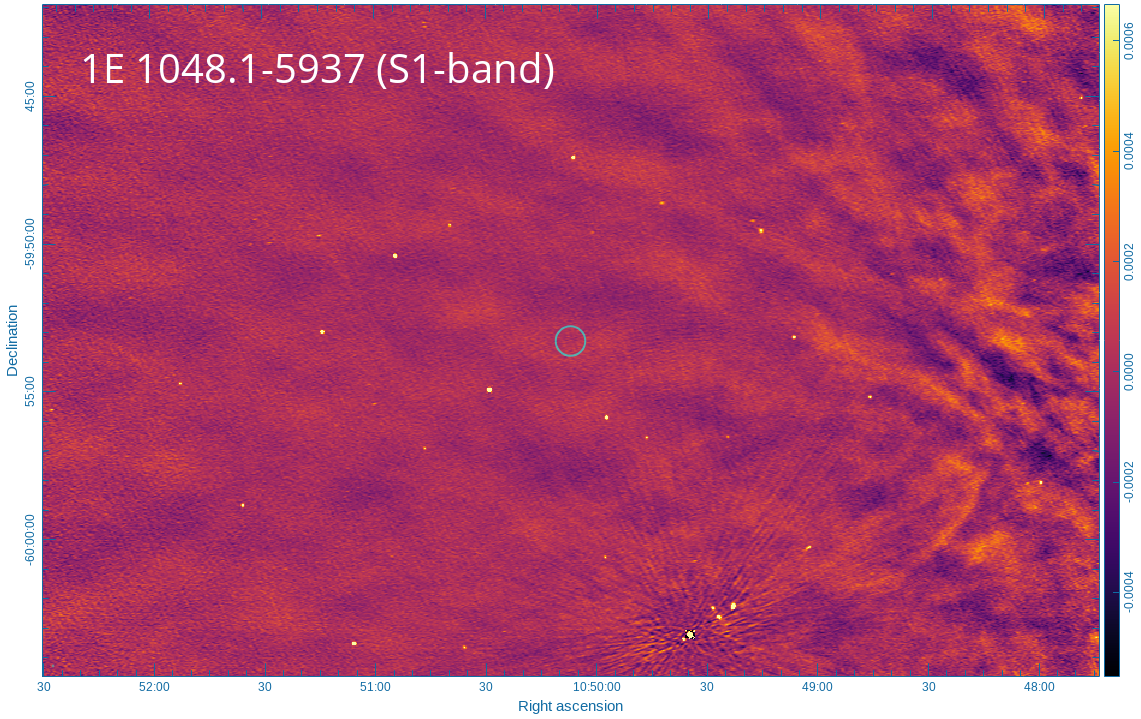}
    \includegraphics[width=0.45\textwidth]{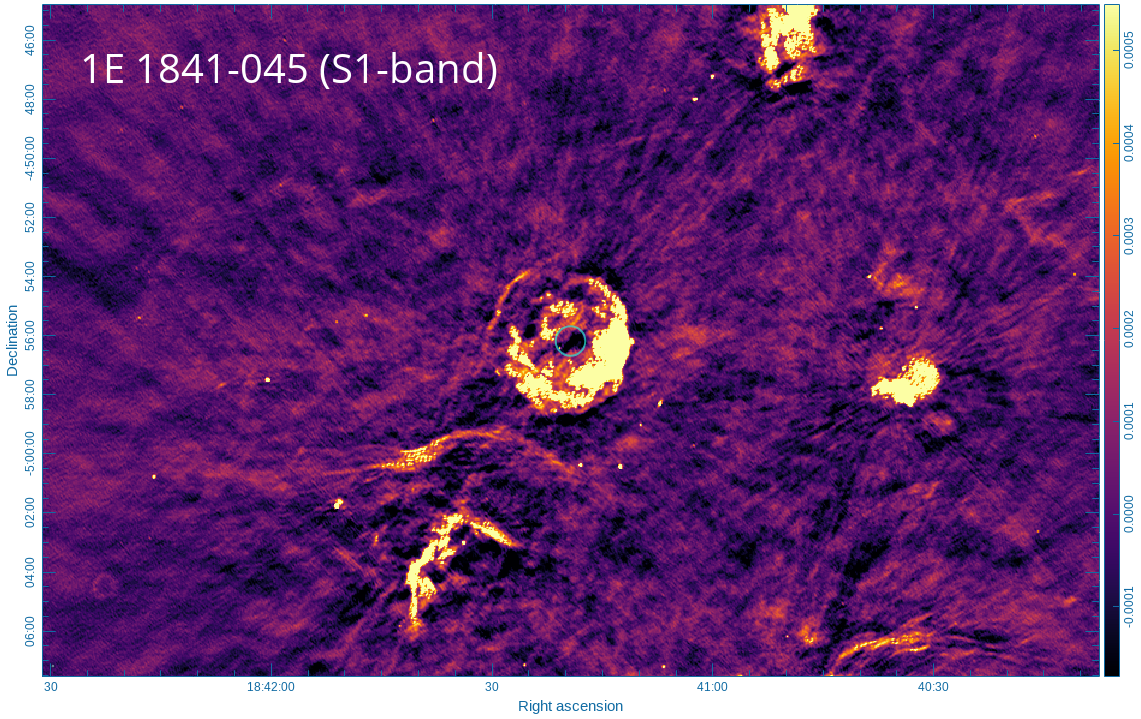}
    \includegraphics[width=0.45\textwidth]{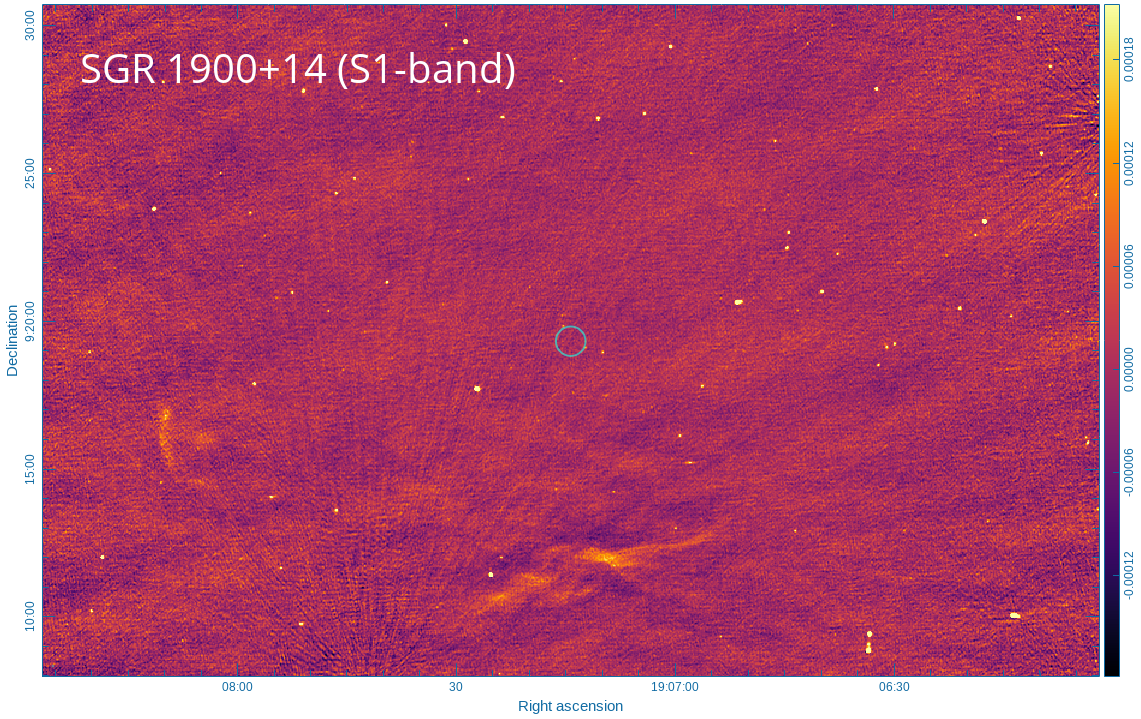}
    \includegraphics[width=0.45\textwidth]{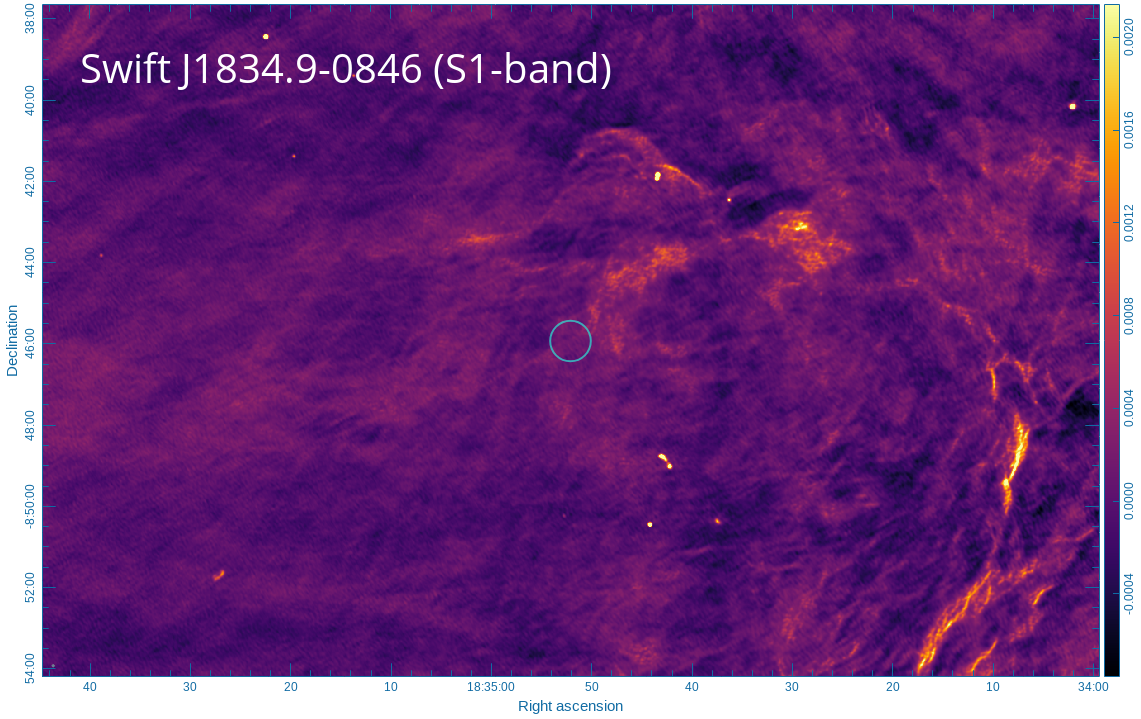}
\label{fig:mkt_sband_all}
\caption{Images of all other sources at S1-band}
\end{figure}

\begin{figure*}
    \centering
    \includegraphics[width=0.45\textwidth]{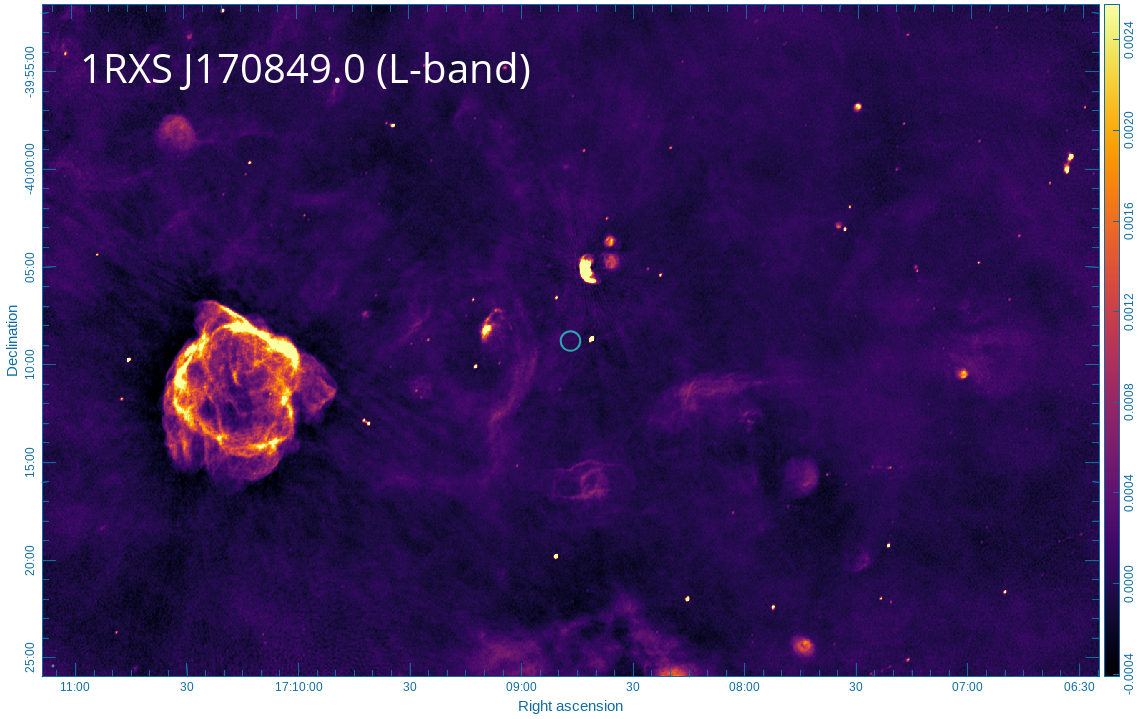}
    \includegraphics[width=0.45\textwidth]{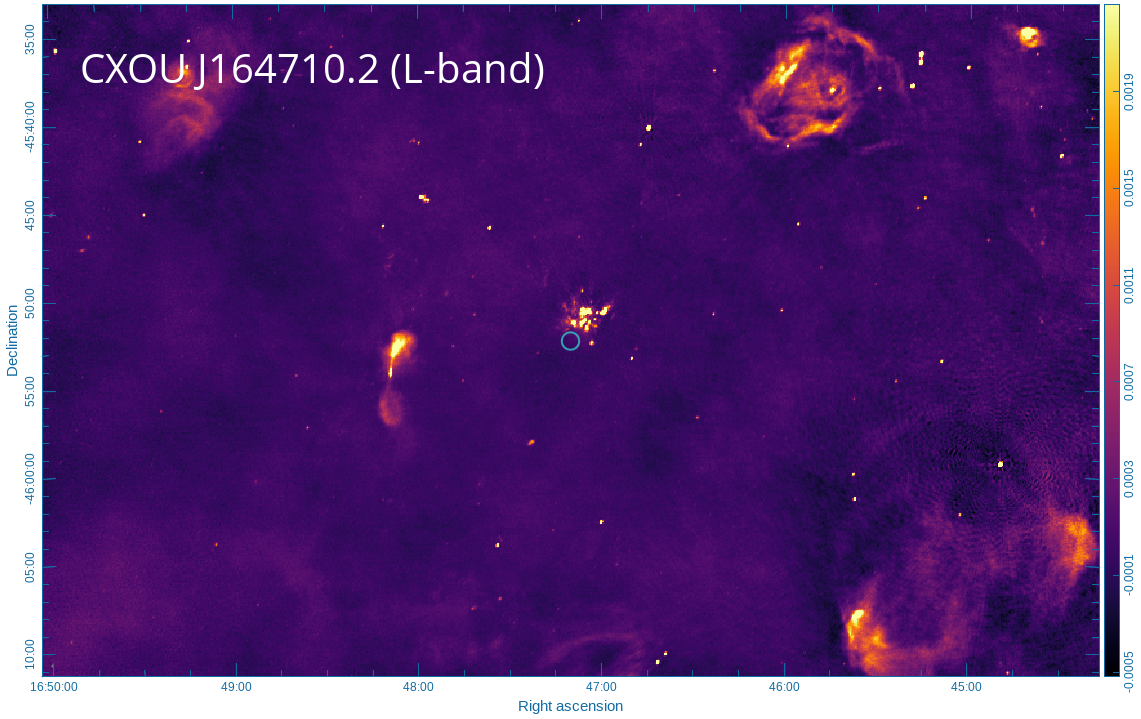}
    \includegraphics[width=0.45\textwidth]{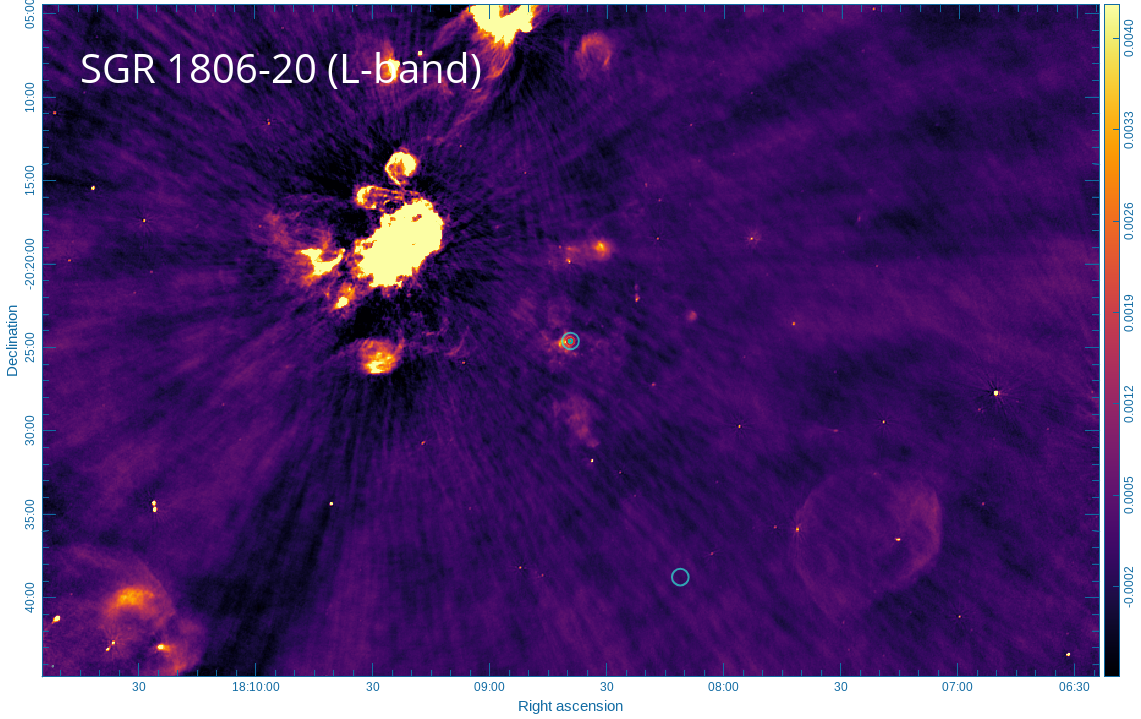}
    \includegraphics[width=0.45\textwidth]{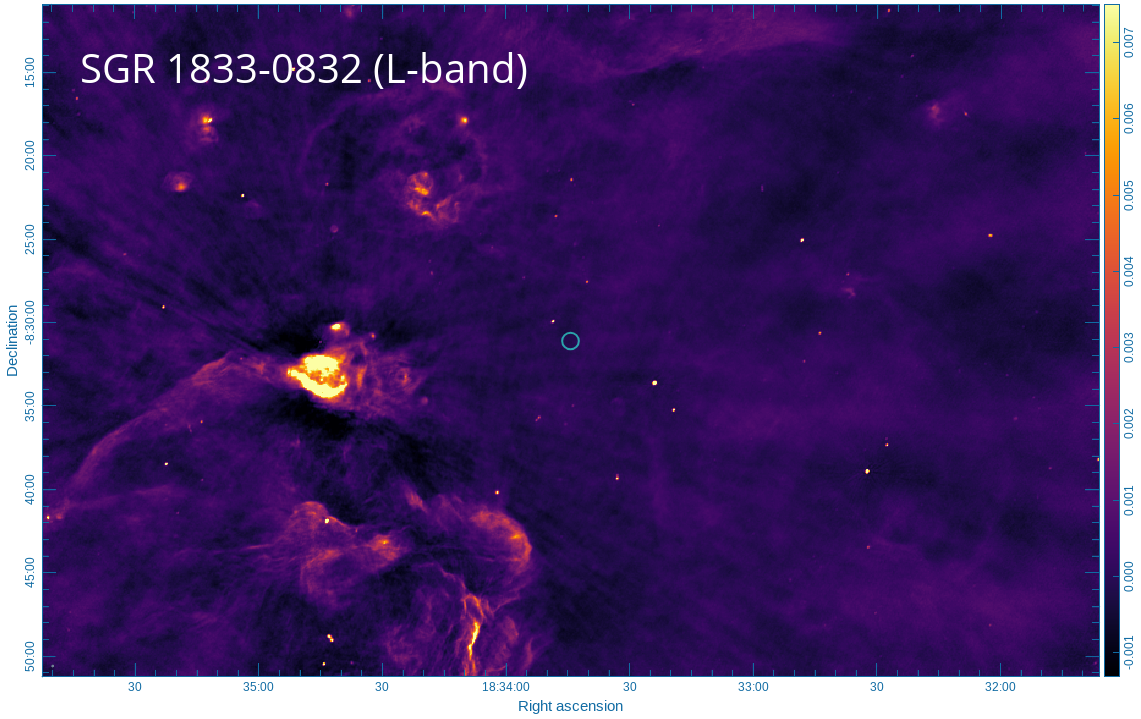}
    \includegraphics[width=0.45\textwidth]{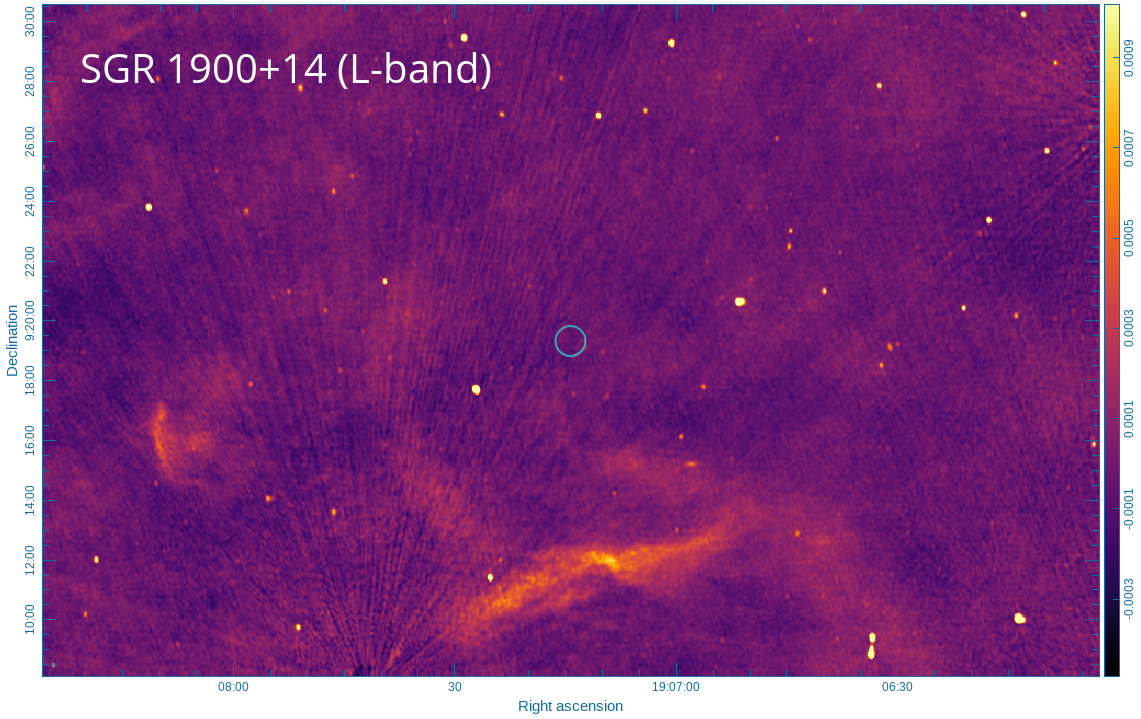}
    \includegraphics[width=0.45\textwidth]{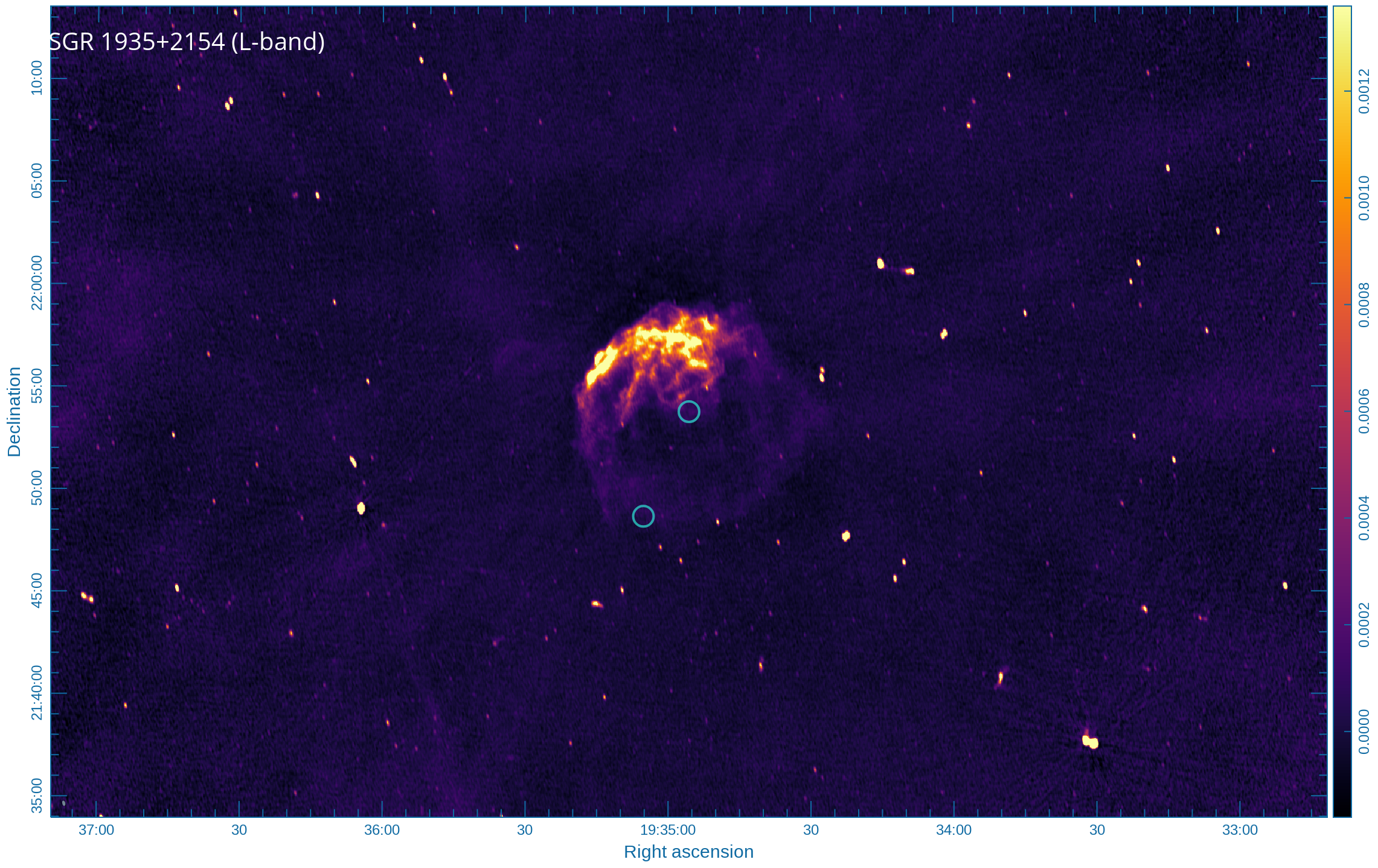}
    \includegraphics[width=0.45\textwidth]{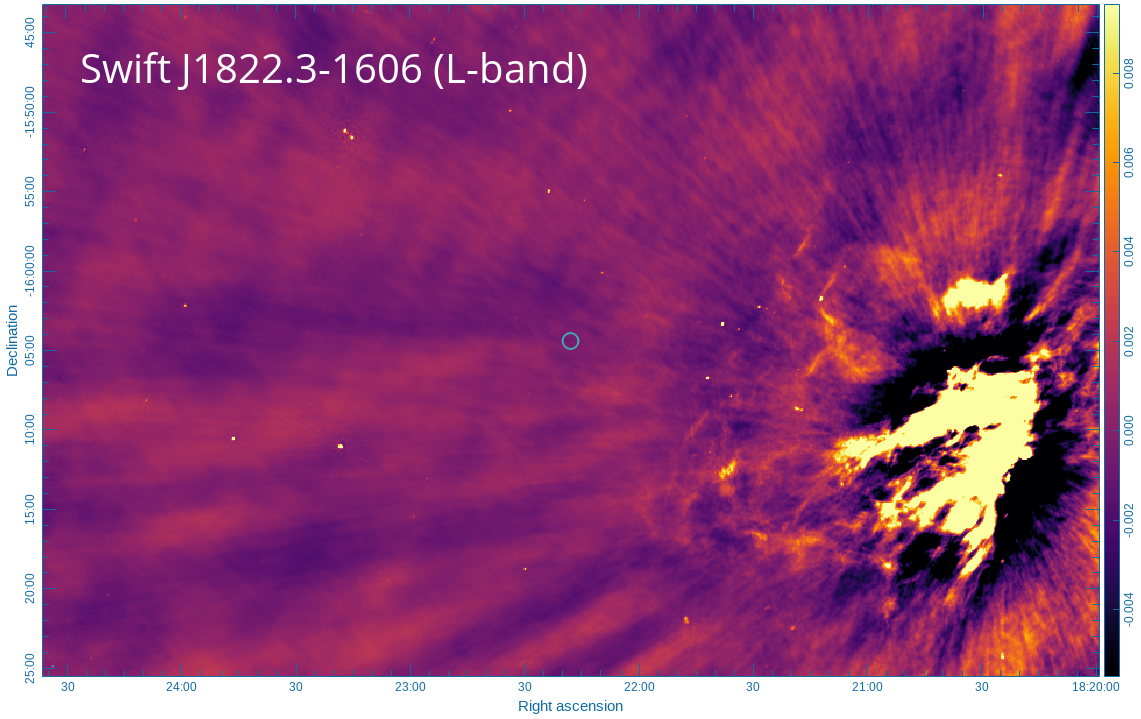}
    \includegraphics[width=0.45\textwidth]{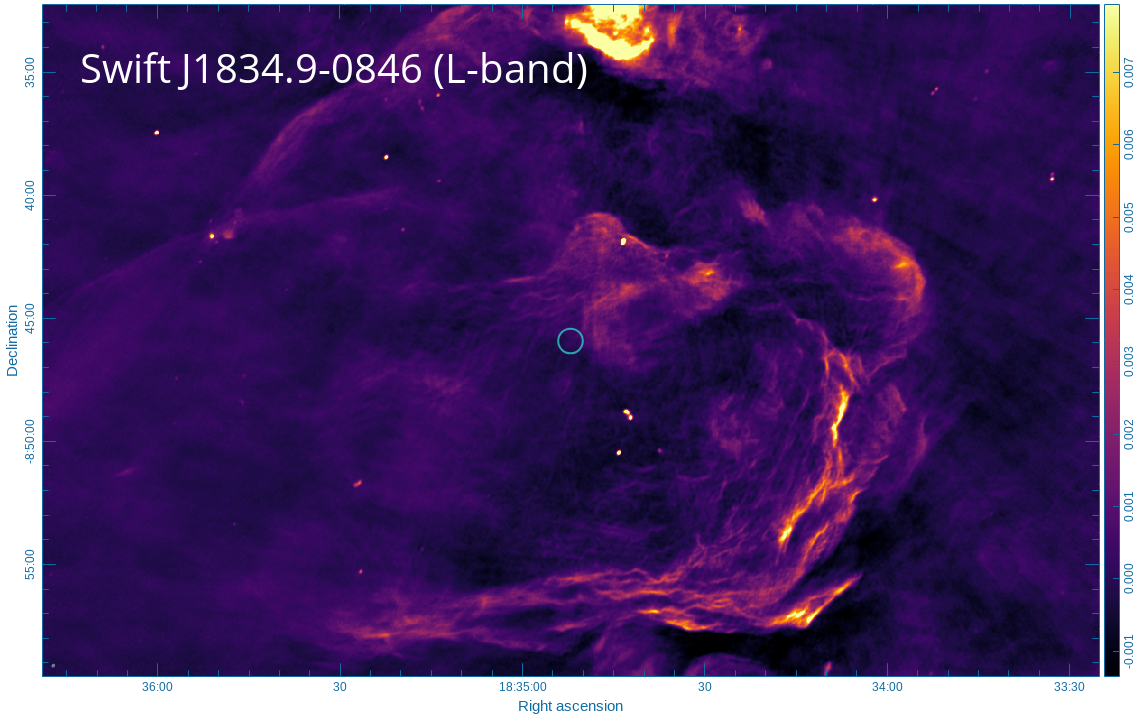}

\label{fig:mkt_lband_all}
\caption{Images of all other sources at L-band}
\end{figure*}

\begin{figure*}
    \centering
    \includegraphics[width=\linewidth]{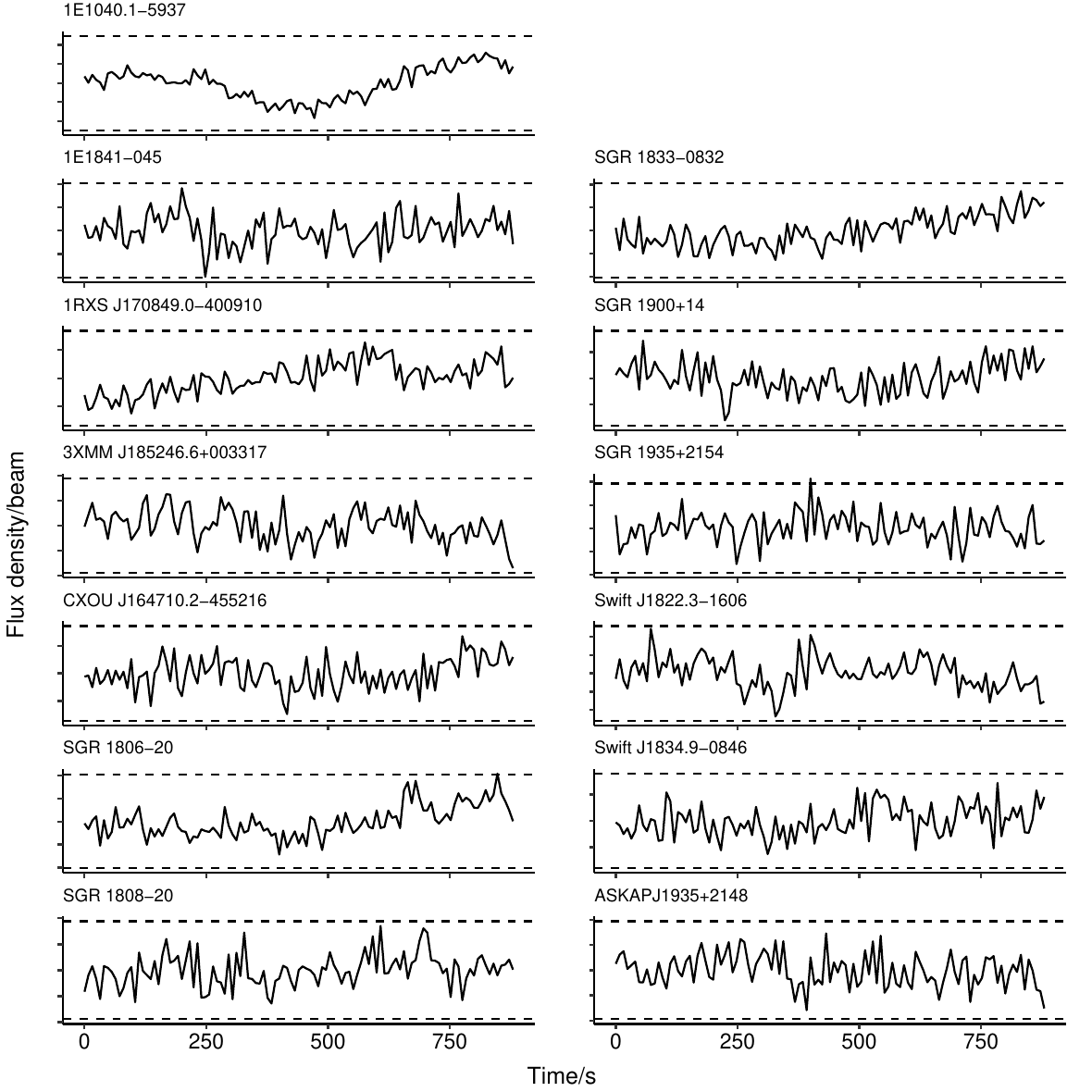}
    \caption{Light curves (solid lines) and +-3 sigma limits (dashed lines) for each source for the L-band observation for the magnetars as well as the ULP ASKAP J1935+2148, which is in the beam of the SGR 1935+2154 observation. The time resolution is \SI{8}{s}.}
    \label{fig:lightcurves_lband}
\end{figure*}

In addition to the images presented in \cref{fig:meerkat_images}, \cref{fig:mkt_sband_all} and \cref{fig:mkt_lband_all} present the final, primary beam corrected images of the magnetars with the positions of the magnetars highlighted.
\Cref{fig:lightcurves_lband} presents the light curves of each magnetar and the ultra long period transient in the image of SGR 1935+2148.
%\subsection{Total intensity images of Epoch 6 (S1-band)}

\begin{figure}
    \centering
\includegraphics[width=0.45\textwidth]{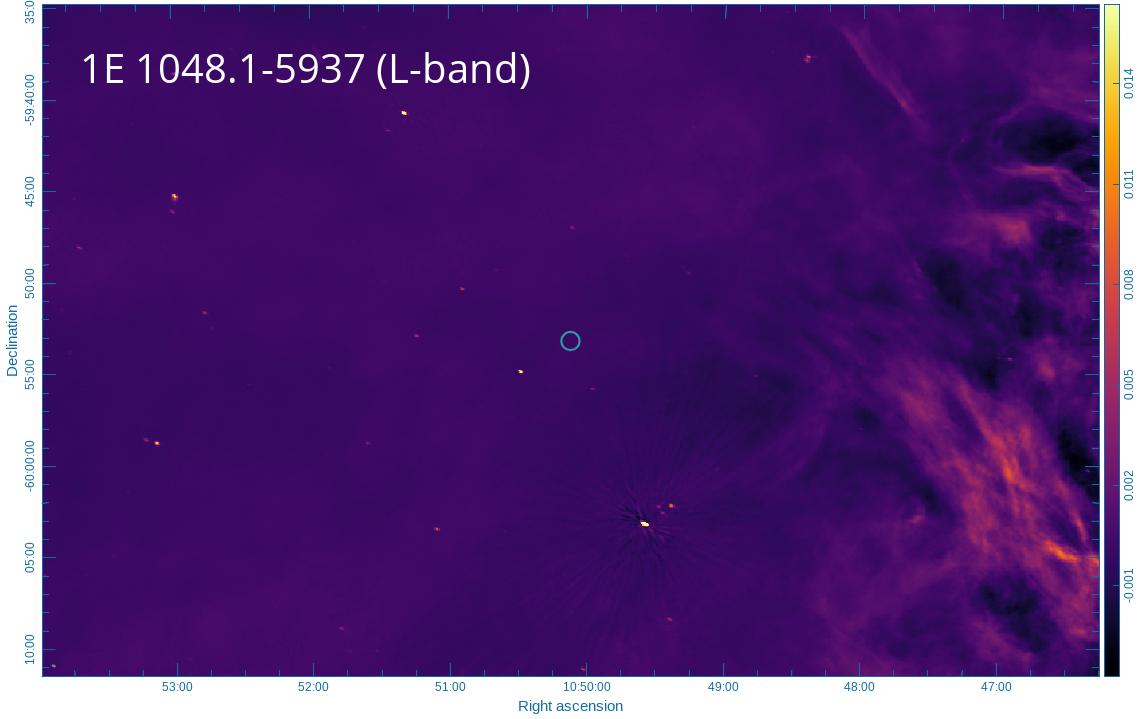}
\includegraphics[width=0.45\textwidth]{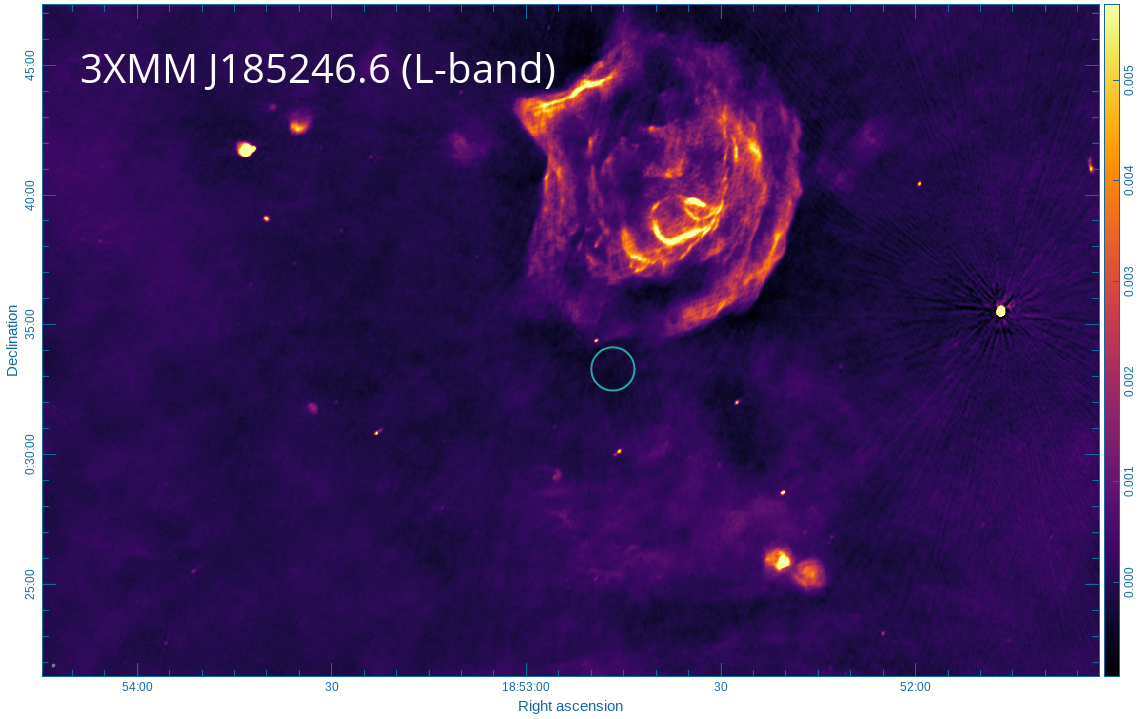}
\label{fig:mkt_band_all_1}
\caption{Images of all other sources at S1-band}
\end{figure}

%\subsection{Total intensity images of Epoch 7 (L-band)}

%\subsection{Light curves of the L-band observation}
\end{appendix}

\end{document}